\documentclass[journal]{IEEEtran}

\usepackage{amsmath}
\usepackage{amsfonts}
\usepackage{amssymb}
\usepackage{amsthm}
\usepackage{tabularx}
\usepackage{mathrsfs} 

\usepackage{amsmath,amssymb,amsfonts}
\usepackage{amsmath,amssymb,amsfonts} 
\usepackage{algorithm}    
\usepackage{algorithmic}  
\usepackage{adjustbox} 
\usepackage{url}
\usepackage[colorlinks]{hyperref}
\newtheorem{theorem}{Theorem}
\newtheorem{remark}{Remark}

\newtheorem{corollary}{Corollary}

\usepackage{stfloats}
\usepackage{float}
\usepackage{graphicx}
\usepackage{xcolor}
\usepackage{subfigure}
\newcolumntype{C}[1]{>{\centering\arraybackslash}p{#1}}
\usepackage{threeparttable}

\makeatletter
\def\blfootnote{\xdef\@thefnmark{}\@footnotetext}
\makeatother

\begin{document}

\title{\huge{Coverage Region Characterization and Optimization of FIRES-Assisted NOMA and OMA Systems}}

\author{Farshad~Rostami~Ghadi,~\IEEEmembership{Senior Member}, \textit{IEEE}, 
Kai-Kit~Wong,~\IEEEmembership{Fellow},~\textit{IEEE}, 
Masoud~Kaveh,~\IEEEmembership{Member},~\textit{IEEE},\\
Hanjiang~Hong,~\IEEEmembership{Member},~\textit{IEEE}, 
Chan-Byoung~Chae,~\IEEEmembership{Fellow},~\textit{IEEE},~and 
Lajos~Hanzo,~\IEEEmembership{Life Fellow},~\textit{IEEE}
\vspace{-5mm}
}
\maketitle

\blfootnote{The work of F. Rostami Ghadi is supported by the European Union's Horizon 2022 Research and Innovation Programme under Marie Sk\l odowska-Curie Grant No. 101107993. The work of K. K. Wong is supported by the Engineering and Physical Sciences Research Council (EPSRC) under Grant EP/W026813/1. The work of C.-B. Chae was supported by NRF/IITP (RS-2024-00397216, RS-2024-00428780) grants funded by the Korean government. 
The work of L. Hanzo is supported by the Engineering and Physical Sciences Research Council (EPSRC) projects Platform for Driving Ultimate Connectivity (TITAN) under Grant EP/X04047X/1 and Grant EP/Y037243/1.}

\blfootnote{\noindent F. Rostami Ghadi is with the Department of Signal Theory, Networking and Communications, Research Centre for Information and Communication Technologies (CITIC-UGR), University of Granada, 18071, Granada, Spain. (e-mail: $\rm f.rostami@ugr.es$).}
\blfootnote{\noindent K. K. Wong and H. Hong are affiliated with the Department of Electronic and Electrical Engineering, University College London, Torrington Place, WC1E 7JE, United Kingdom. K. K. Wong is also affiliated with Yonsei Frontier Lab, Yonsei University, Seoul, Korea. (e-mail: $\rm \{kai\text{-}kit.wong, hanjiang.hong\}@ucl.ac.uk$).}


\blfootnote{\noindent M. Kaveh is with the  Department of Information and Communication Engineering, Aalto University, Espoo, Finland. (e-mail: $\rm masoud.kaveh@aalto.fi$).}
\blfootnote{\noindent C-B. Chae is with the School of Integrated Technology, Yonsei University, Seoul, Korea (e-mail: $\rm cbchae@yonsei.ac.kr$).}
	

\blfootnote{\noindent L. Hanzo is with the School of Electronics and Computer Science, University of Southampton, Southampton, U.K. (e-mail: $\rm lh@ecs.soton.ac.uk$).}
\blfootnote{Corresponding author: Kai-Kit Wong.}

\begin{abstract}
This paper investigates the fluid integrated reflecting and emitting surfaces (FIRES)-aided communication system. In these metasurfaces, each subarea hosts an active element capable of simultaneous transmission and reflection, phase, and geometric positioning control within the subarea. We develop a coverage-centric system model for the two-user downlink scenario (one user per half-space) under spatially correlated Rician fading and imperfect phase control. First, we derive closed-form far-field line-of-sight (LoS) coverage bounds that reveal the effects of aperture size, base station (BS) distance, transmit power, energy splitting (ES), and phase errors. Protocol-aware corollaries are then presented for both orthogonal multiple access (OMA) and non-orthogonal multiple access (NOMA), including conditions for successful successive interference cancellation (SIC). Second, we formulate coverage maximization as a bi-level optimization problem consisting of (i) an outer search over FIRES element positions, selecting one active preset per subarea under minimum-spacing constraints, and (ii) an inner resource allocation problem tailored to the multiple-access scheme, which is one-dimensional for OMA and a low-dimensional numerical search for NOMA. The proposed framework explicitly accounts for target rate constraints, ES conservation, power budgets, geometric placement limits, and decoding-order feasibility. Extensive simulations demonstrate that FIRES, by jointly exploiting geometric repositioning and passive energy control, substantially enlarges the coverage region compared with a conventional simultaneously transmitting and reflecting reconfigurable intelligent surface (STAR-RIS) under the same element budget. Furthermore, NOMA yields additional coverage gains when feasible. The analytical coverage bounds closely match the simulation results and quantify the robustness of FIRES to phase-control imperfections.
\end{abstract}


\vspace{-2mm}
\section{Introduction}\label{sec-intro}
\IEEEPARstart{T}{he proliferation}  of Internet of Things (IoT) networks has created increasingly stringent requirements for reliable and wide-area wireless coverage, particularly in environments with dense obstacles, heterogeneous device locations, and severely blocked propagation paths \cite{ngu20226g,gu2021en}. Such coverage limitations are especially relevant in smart factories, warehouses, smart buildings, and other large-scale IoT deployments, in which distributed devices need to maintain reliable connectivity despite unfavorable propagation conditions. In this context, the evolution toward next-generation wireless networks has spurred the development of reconfigurable intelligent surfaces (RISs) as an enabling technology for smart radio environments \cite{basar2019,renzo2020}. By controlling the electromagnetic response of large arrays of nearly passive elements \cite{pan2021ris}, RISs enable programmable propagation that can enhance spectral efficiency, coverage, and energy performance at less hardware cost. However, the conventional RIS paradigm remains limited to unidirectional reflection, restricting its coverage to a single half-space and confining the available design degrees of freedom (DoF) primarily to phase control. This limitation can be particularly restrictive in IoT environments where devices are distributed across different spatial regions and reliable connectivity is required beyond one side of the surface.

To overcome this limitation, the idea of simultaneously transmitting and reflecting RISs (STAR-RISs) was introduced \cite{liu2021star}. In a STAR-RIS, each unit element is capable of splitting the incident energy into two independent parts for transmission and reflection, controlled via amplitude and phase coefficients. This innovation extends the RIS coverage to the entire $360^\circ$ spatial domain, thereby enabling full-space communications in which users on both sides of the surface can be served simultaneously. STAR-RISs have proven effective in enlarging coverage regions and improving rate performance, especially when integrated with multiple access techniques such as non-orthogonal multiple access (NOMA) and orthogonal multiple access (OMA) \cite{yue2023star}. Nevertheless, existing STAR-RIS implementations remain geometrically static, as their meta-atoms are fixed in position, which constrains the attainable DoF for spatial adaptation in dynamic wireless environments.\footnote{Furthermore, stacked intelligent metasurfaces (SIMs) have more recently emerged as another promising paradigm for enhancing the electromagnetic processing capabilities of programmable surfaces. By cascading multiple configurable metasurface layers, SIMs enable wave domain signal processing through the joint control of inter-layer propagation and phase responses \cite{zhnag2026skat,wei2026near,q2025}. SIM architectures primarily exploit the additional processing DoF offered by multiple stacked layers, but they are unsuitable for RIS settings because SIMs are bulky and computationally complex.}

At the same time, with the seminal work by Wong {\em et al.}~in \cite{wong2020per,wong2021fluid}, fluid antenna systems (FASs) have emerged as a hardware-agnostic system concept that considers the antenna as a reconfigurable physical-layer resource to broaden system design and network optimization \cite{new2025tut,Lu-2025,hong2025contemporary,new2025flar,wu2024flu}. FAS focuses on using antenna shape and position flexibility by reconfiguring electromagnetic parameters via movable elements \cite{zhu2024historical}, liquid-based structures \cite{shen2024design,Shamim-2025}, metamaterials \cite{Zhang-jsac2026,Liu-2025arxiv}, or reconfigurable pixels \cite{zhang2024pixel,tong-2025pixel,Wong-wc2026,qiao-2026,zhang-2026pixel}, etc. Through antenna reconfiguration, FAS can exploit spatial diversity fully over a given spatial aperture with less and even only a single radio frequency (RF) chain, offering substantial gains in outage performance, and interference suppression. 

Following its introduction, interest in FAS has grown considerably. Initial efforts have focused on studying the diversity order under rich scattering channels \cite{New2023fluid,new2024inf}. Recent progress has also been made to understand the performance of FAS in the finite blocklength regime \cite{Zhang-2026fas}. Furthermore, results have shown great benefits of using FAS on existing multiple access technologies \cite{new2024noma,ghadi2025fluid}. Wireless power transfer and integrated sensing and communication (ISAC) systems have also found FAS particularly beneficial due to the additional spatial DoF \cite{ghadi2024paris,mao2025joint,zou2024isac,ghadi2025isac,zh2024isac,Wang-2024isac}. Besides, FAS has been applied in physical layer security (PLS) \cite{tang2023fluid,rostami2024phy}, unmanned aerial vehicle (UAV) systems \cite{ghadi2025uav,jiang2025dy}. More fundamentally, FAS enables a novel way for multiple access, referred to as fluid antenna multiple access (FAMA) \cite{{Hong-fama2026}}. Slow FAMA \cite{wong2023sFAMA}, fast FAMA \cite{wong2022FAMA,wong2023fast}, compact ultra-massive array (CUMA) \cite{wong2024compact} and coded FAMA \cite{waqar2025turbocharging,hong2025Downlink} are some of the emerging multiple access techniques, facilitated by FAS. Integration of FAS with multicarrier systems \cite{hong2025fas} and channel estimation for FAS \cite{xu2024channel} have also been investigated and tackled recently.

\vspace{-2mm}
\subsection{FAS and RIS Synergy}
The success of FASs has inspired their integration into RIS technology \cite{ghadi2024per}, giving rise to the concept of the fluid reconfigurable intelligent surface (FRIS) \cite{salem2025first}. A FRIS generalizes the conventional RIS by introducing position reconfigurability at the element level. Precisely, each element of FRIS, referred to as a ``fluid'' element, can change its position within a designated subarea of the surface while simultaneously adjusting its phase and/or amplitude responses. This unique capability enables FRIS to reshape not only the electromagnetic phase front but also the geometrical distribution of its scattering points, leading to additional spatial DoF beyond those provided by conventional RIS or even STAR-RIS architectures. By adaptively positioning its elements according to the channel conditions, FRIS can maximize the received power, mitigate fading, and enhance link robustness without increasing the total number of elements. Early analytical studies have revealed that FRIS-assisted systems can achieve substantial improvements in ergodic capacity and outage probability performance compared to traditional fixed RISs, primarily owing to this dynamic spatial adaptability \cite{xiao2025fluid,ghadi2025perf,kaveh2025pls,xiao2025ffris}.

Furthering these advances, the fluid integrated reflecting and emitting surface (FIRES) concept has recently been proposed as an extension to FRIS and STAR-RIS and a combination of both for wireless communication \cite{ghadi2025fires}. FIRES merges the dual functionality of the STAR-RIS with the spatial reconfigurabiity of the FRIS into a single unified architecture. Unlike conventional FAS, where spatial reconfigurability is realized at the transceiver antenna, and conventional RIS/STAR-RIS architectures, where the metasurface elements remain spatially fixed, FIRES introduces element-enabled spatial reconfigurability directly into a full-space transmitting and reflecting metasurface. 
Each fluid unit-cell in FIRES is an active subarea element capable of simultaneously performing transmission and reflection, controlling the per-side phase response, and repositioning within its subarea in almost real time. In effect, FIRES introduces a triple-tunable design space, namely position, phase, and power-splitting control, that jointly governs the propagation environment. Unlike STAR-RISs, which can only adjust the electromagnetic parameters, FIRES leverages both electromagnetic and geometric reconfigurability, creating a \textit{fluid} aperture whose spatial pattern can dynamically adapt to user locations and channel evolution. This renders FIRES a  versatile platform for full-space wireless coverage, beam manipulation, and interference management.

\vspace{-2mm}
\subsection{State-of-the-Art}
Though the concept of FRIS is in an early stage, it represents a natural progression in the evolution of programmable wireless environments \cite{xiao2025from}. Early efforts, such as \cite{salem2025first}, introduced FRIS as an extension to the conventional RISs, in which each reflecting element can be physically repositioned within a designated subarea of the surface, thereby increasing the spatial DoF. By jointly optimizing the element positions, phase shifts, and base station (BS) precoding in both single-user and multiuser systems, FRIS was demonstrated to achieve substantial rate improvements over static RIS architectures. Building upon this foundation, \cite{xiao2025fluid} proposed a more practical discrete activation model for FRIS, where each fluid element is realized through a dense matrix of controllable sub-elements. Specific sub-elements are dynamically activated and phase-tuned according to instantaneous channel conditions. Employing a cross-entropy optimization (CEO) framework, this design achieved notable throughput gains while maintaining reduced implementation complexity. While these studies demonstrated the performance benefits of element-wise position reconfigurability and provided practical mechanisms for discrete activation, their focus remained primarily on rate optimization in reflection-based FRIS architectures, and full-space coverage by FRIS or FIRES is still not well understood.

To further establish the theoretical underpinnings of FRIS, \cite{ghadi2025perf} presented an analytical performance characterization by deriving closed-form statistical expressions for the equivalent end-to-end channel. The authors obtained approximations for both the outage probability and ergodic capacity, including their asymptotic behaviors, revealing key performance trends that confirm the advantage of dynamic element activation in improving link reliability and spectral efficiency compared with RISs. Furthermore, \cite{kaveh2025pls} explored PLS of FRIS-assisted wireless systems, where a BS communicates with a legitimate user with an eavesdropper. By dynamically activating a subset of elements to adapt to channel conditions, FRIS enhances spatial diversity and secrecy capacity. Analytical bounds for the secrecy outage probability and average secrecy capacity were derived, showing that even with partial activation, FRIS significantly outperforms conventional RISs in secure communications.  These analytical studies provided valuable insights into the reliability and security benefits of FRIS under fading and spatial correlation. However, they considered performance metrics such as outage, ergodic capacity, and secrecy rather than the achievable coverage region, and did not incorporate simultaneous transmission and reflection.

On the other hand, \cite{xiao2025ffris} proposed a pattern-reconfigurable FRIS. Compared to position-reconfigurable and conventional RISs, this design achieved superior signal enhancement. This work extended to a multiuser system, jointly optimizing beamforming and spherical harmonics coefficients via a minimum mean-square error (MMSE) approach with the aid of Riemannian conjugate gradient (RCG) algorithm. The results showed performance gains exceeding $160\%$ over traditional RISs, demonstrating the potential of radiation-pattern reconfigurability. Although pattern reconfigurability introduces an additional means of adapting the surface response and has demonstrated substantial multiuser gains, it represents a very different form of reconfigurability from the element-level geometric repositioning considered in FIRES and does not address full-space coverage optimization. The concept of FIRES was eventually introduced in \cite{ghadi2025fires}, where the authors investigated an FIRES-assisted system under the energy splitting (ES) protocol, in which each subarea of the surface functions as a fluid element capable of simultaneous transmission, reflection, and position adjustment. An optimization problem was formulated to jointly design the position, phase, and power splitting parameters of the surface, and a particle swarm optimization (PSO) algorithm was developed to obtain an efficient solution, where their results again illustrated that FIRES significantly outperforms conventional STAR-RIS architectures in terms of achievable performance. The initial FIRES study established the benefits of jointly controlling element positions, phases, and ES and demonstrated clear gains over conventional STAR-RIS architectures. Nevertheless, it focused primarily on performance optimization and did not provide an analytical characterization or optimization of the bilateral coverage region, nor did it investigate the comparative behavior of OMA and NOMA under practical phase control impairments.

\begin{table*}[ht]
\caption{Comparison of our contributions to the literature}\label{table1}
\centering
\scriptsize
\resizebox{.95\linewidth}{!}{
\begin{tabular}{|C{1.6cm}|C{1.8cm}|C{1.8cm}|C{2.5cm}|C{2.5cm}|C{2.5cm}|C{2.4cm}|}
\hline
\textbf{Works} & \textbf{Fixed element} & \textbf{Fluid element} & \textbf{Coverage optimization} & \textbf{Coverage performance} & \textbf{Half-space coverage} & \textbf{Full-space coverage} \\
\hline
\cite{salem2025first,xiao2025fluid,ghadi2025perf,kaveh2025pls,xiao2025ffris} & $\times$ & $\checkmark$ & $\times$ & $\times$ & $\checkmark$ & $\times$ \\ \hline
\cite{ghadi2025fires} & $\times$ & $\checkmark$ & $\times$ & $\times$ & $\checkmark$ & $\checkmark$ \\ \hline
\cite{wu2021cov} & $\checkmark$ & $\times$ & $\checkmark$ & $\times$ & $\checkmark$ & $\checkmark$ \\ \hline
\cite{ghadi2023cov} & $\checkmark$ & $\times$ & $\times$ & $\checkmark$ & $\checkmark$ & $\checkmark$ \\ \hline
\textbf{Proposed} & $\times$ & $\checkmark$ & $\checkmark$ & $\checkmark$ & $\checkmark$ & $\checkmark$ \\ \hline
\end{tabular}}
\end{table*}

\vspace{-2mm}
\subsection{Motivation and Contributions} 
Despite the promising advances in FRIS and the introduction of FIRES, several important gaps remain. Existing FRIS studies have mainly focused on rate, outage, capacity, secrecy, or beamforming performance in reflection-based or half-space configurations, while the initial FIRES work established the benefit of jointly controlling position, phase, and ES without providing an analytical characterization of the achievable bilateral coverage region. Although previous STAR-RIS studies have studied coverage extension in full-space communications \cite{wu2021cov,ghadi2023cov}, their results cannot be applied to FIRES because of the additional spatial reconfigurability of the surface elements. Three issues remain unaddressed. First, the coverage behavior of FIRES-assisted networks has not been analytically characterized. Second, the joint impact of element repositioning, ES, phase control impairments, and multiple access strategy on the coverage region has not been quantified. Third, the integration of OMA and NOMA with FIRES introduces coupled spatial and resource allocation problems whose coverage implications remain unexplored. To address these, this paper presents a comprehensive coverage analysis and optimization framework for FIRES-assisted NOMA and OMA communication systems. For clarity, Table \ref{table1} positions the proposed framework against the related FRIS, FIRES, and STAR-RIS studies, highlighting the differences in spatial reconfigurability, coverage characterization and optimization, and half-space/full-space operation.

Our main contributions are summarized as follows:
\begin{itemize}
\item \textit{Unified System Model for FIRES-Aided Full-Space Networks:} We develop a general model in which a BS communicates with two users located on opposite sides of a FIRES. Each element simultaneously performs reflection and transmission while adjusting its position within a designated subarea. This model incorporates Rician fading, phase errors, and power-splitting coefficients, capturing the realistic characteristics of fluid metasurfaces.
\item \textit{Analytical Coverage Characterization:} We derive closed-form expressions and tractable approximations for the coverage radius under both OMA and NOMA schemes. The analysis explicitly accounts for spatial reconfigurability, element density, and ES ratios, quantifying how the FIRES parameters affect the aggregate bilateral coverage region.  
\item \textit{Bi-Level Optimization Framework for Coverage Maximization:} To fully exploit the capabilities of FIRES, we formulate a bi-level optimization problem where the outer layer obtains the optimal element positions within each subarea, and the inner layer allocates protocol-specific resources, namely the time fraction for OMA and the power and ES coefficients for NOMA. Efficient schemes are developed for both OMA and NOMA scenarios: a one-dimensional (1D) search over the time fraction for the former and a low-dimensional numerical search over the power and ES variables for the latter.
\item \textit{Comparative Evaluation and Insights:}
Extensive simulations validate the accuracy of the analytical derivations and demonstrate that FIRES provides a substantial coverage gain over conventional STAR-RIS architectures. The results further reveal the existence of trade-offs among the number of fluid elements, their spatial positioning range, and the power-splitting factor, offering valuable design guidelines for future FIRES implementations.
\end{itemize}

Therefore, our work bridges the gap between metasurface-level `fluid' reconfigurability (electromagnetic and geometric) and network-level coverage optimization, establishing the first accurate analytical framework for FIRES-assisted NOMA and OMA systems. Our analysis and design principles can serve as a foundation for broader applications of fluid metasurfaces in full-space future communication networks.

\vspace{-2mm}
\subsection{Notations}\label{sec:notation}
Lowercase, bold lowercase, and bold uppercase letters denote scalars, vectors, and matrices, respectively. The operators $(\cdot)^{\mathsf{T}}$ and $(\cdot)^{\mathsf{H}}$ denote the transpose and Hermitian transpose, respectively. Also, $\mathrm{diag}(\mathbf{a})$ is a diagonal matrix whose diagonal entries are taken from vector $\mathbf{a}$, while $\mathrm{Diag}(\mathbf{A})$ extracts the diagonal elements of matrix $\mathbf{A}$ into a vector. $\|\cdot\|_2$ and $\|\cdot\|_{\mathrm{F}}$ are the Euclidean and Frobenius norms, respectively. $\angle z$ and $|z|$ denote the phase and magnitude of a complex number $z$. $\mathbb{E}[\cdot]$ denotes statistical expectation. $\mathbb{C}^{m\times n}$ and $\mathbb{R}^{m\times n}$ denote the sets of $m\times n$ complex and real matrices, respectively. The notation $\mathbf{I}_N$ is an $N\times N$ identity matrix, and $\mathbf{0}_N$ is an $N\times 1$ all-zero vector. Unless otherwise stated, all logarithms are to base~2, and all random variables are assumed to be circularly symmetric complex Gaussian (CSCG) when applicable. For clarity, a summary of the main symbols and system parameters used throughout the paper is given in Table~\ref{table:symbols}.

\vspace{-2mm}
\subsection{Paper Organization}
The rest of this paper is organized as follows. In Section~\ref{sec:system}, we introduce the system and signal models for the FIRES-assisted downlink NOMA and OMA systems, including the channel modeling, signal formulation, and coverage definitions. Then Section~\ref{sec:coverage} provides the analytical characterization of the coverage radius for both multiple-access schemes under Rician fading. Section~\ref{sec:max_coverage} presents the coverage maximization problems for OMA and NOMA, in which the inner problem for OMA simplifies to a 1D search over the time fraction, while the NOMA case is handled through a low-dimensional resource allocation search with adaptive selection between the two candidate decoding orders. Section~\ref{sec:pos_opt} develops the position-aware optimization framework based on a bi-level PSO approach, in which the outer loop searches over discrete element positions subject to spacing and one-active-preset constraints and the inner loop solves the OMA or NOMA coverage problem for each candidate configuration. Section~\ref{sec:sim} presents numerical results and discussions, validating the analytical models and demonstrating the coverage advantages of FIRES over conventional STAR-RIS configuration. Finally, Section~\ref{sec:conclusion} concludes the paper and highlights promising future research directions.

\begin{table}[!t]
\caption{Summary of Key Notations and Parameters}\label{table:symbols}
\centering
\resizebox{.95\linewidth}{!}{
\begin{tabular}{ll}
			\hline
			\textbf{Symbol} & \textbf{Description} \\
			\hline
			$u\in\left\{r,t\right\}$ & User/side index: reflection ($r$) and transmission ($t$)  \\
			$\zeta\in\left\{\mathrm{O,N}\right\}$ & Multiple access index: OMA ($\mathrm{O}$) and NOMA ($\mathrm{N}$)  \\
			$M$ & Number of fluid elements \\
			$N_m$ & Number of preset positions \\
			$\mathbf{r}_m$ & Preset position of $m$-th fluid elements \\
			 $A$ & Physical aperture area \\
			$d_f$ & Distance between BS and FIRES \\
			$D_u$ & Distance between FIRES and user $u$\\
			$D_\mathrm{tot}$ & Aggregate bilateral coverage extent\\
			$D$ & Minimum spacing distance between fluid elements\\
			$f_c$ &  Carrier frequency \\
			$\lambda$ & Wavelength \\
			$\mathbf{h}_f$ & BS-to-FIRES channel vector \\
			$\mathbf{h}_u$ & FIRES-to-user $u$ channel vector \\
			$l_q$ & Large-scale loss for hop $q\in\left\{f,r,t\right\}$ \\
			$K_q$ & Rician factor for hop $q\in\left\{f,r,t\right\}$\\
			$\mathbf{a}_f(\cdot)$ &  Steering vector for BS-to-FIRES \\
			$\mathbf{a}_u(\cdot)$ & Steering vector for FIRES-to-user $u$ \\
			$\mathbf h_{q,{\mathrm{LoS}}}$ & LoS component\\
			$\mathbf h_{q,{\mathrm{NLoS}}}$ & NLoS component\\
			$\mathbf{\Lambda}_q$ & Diagonal matrix of eigenvalues \\
			$\overline{\mathbf{h}}_q$  & Uncorrelated small-scale fading component\\
			$\mathbf{R}_q$ & Spatial correlation matrix\\
			$\mathcal{F}(\cdot)$ & Mapping function from 2D port indices to 1D index \\
			$\beta_u$ & Power-splitting coefficient \\
			$\alpha$ & Path-loss exponent \\
			$\rho_0$ & Unit-distance power gain \\
			$\mathbf{\Phi}_u$ & Per-side passive transfer matrix\\
			$\phi_m^u$ & Programmed phase\\
			$\epsilon_m^u$ & residual phase error\\
			$\theta_m^u$ & Per-cell cascaded phase\\
			$h_\mathrm{eff}^u$ & Effective cascaded channel\\
			$\chi_u$ & Phase-error attenuation factor\\
			$H_u(\mathbf r)$ & Phase-aligned cascaded gain\\
			$x$  & Transmit symbol\\
			$P$ & Transmit power\\
			$z_u$ & AWGN noise at user $u$\\
			$\gamma_u$ & Instantaneous SNR at user $u$\\
			$R_u$ & ES Rate\\
			$\tau_u$ & OMA resource fraction\\
			$R_u^{\rm tar}$ & Target rate\\
			$\gamma_{\mathrm{th},u}$ & SNR threshold\\
			$p_u$ & NOMA BS power fractions\\
			$\widetilde G_u(\mathbf r)$ & Effective link gain used by inner programs\\
			$\kappa$ & Link-disparity factor \\
			$N_p$ & Swarm size of PSO \\
			$T$ & Iterations of PSO\\ 
			$w$ & Inertia weight in PSO updates\\
			$c_1$ & Cognitive acceleration coefficient\\
			$c_2$ & Social acceleration coefficient \\
			$v_{\max}$ & Velocity clamp per dimension. \\
			$N_h^{\rm sub}$ & Horizontal discrete preset resolution inside each subarea\\ $N_v^{\rm sub}$ & Vertical discrete preset resolution inside each subarea\\
\hline
\end{tabular}}
\end{table}

\vspace{-2mm}
\section{System  Model}\label{sec:system}
\begin{figure}[!t]
\centering
\includegraphics[width=0.9\columnwidth]{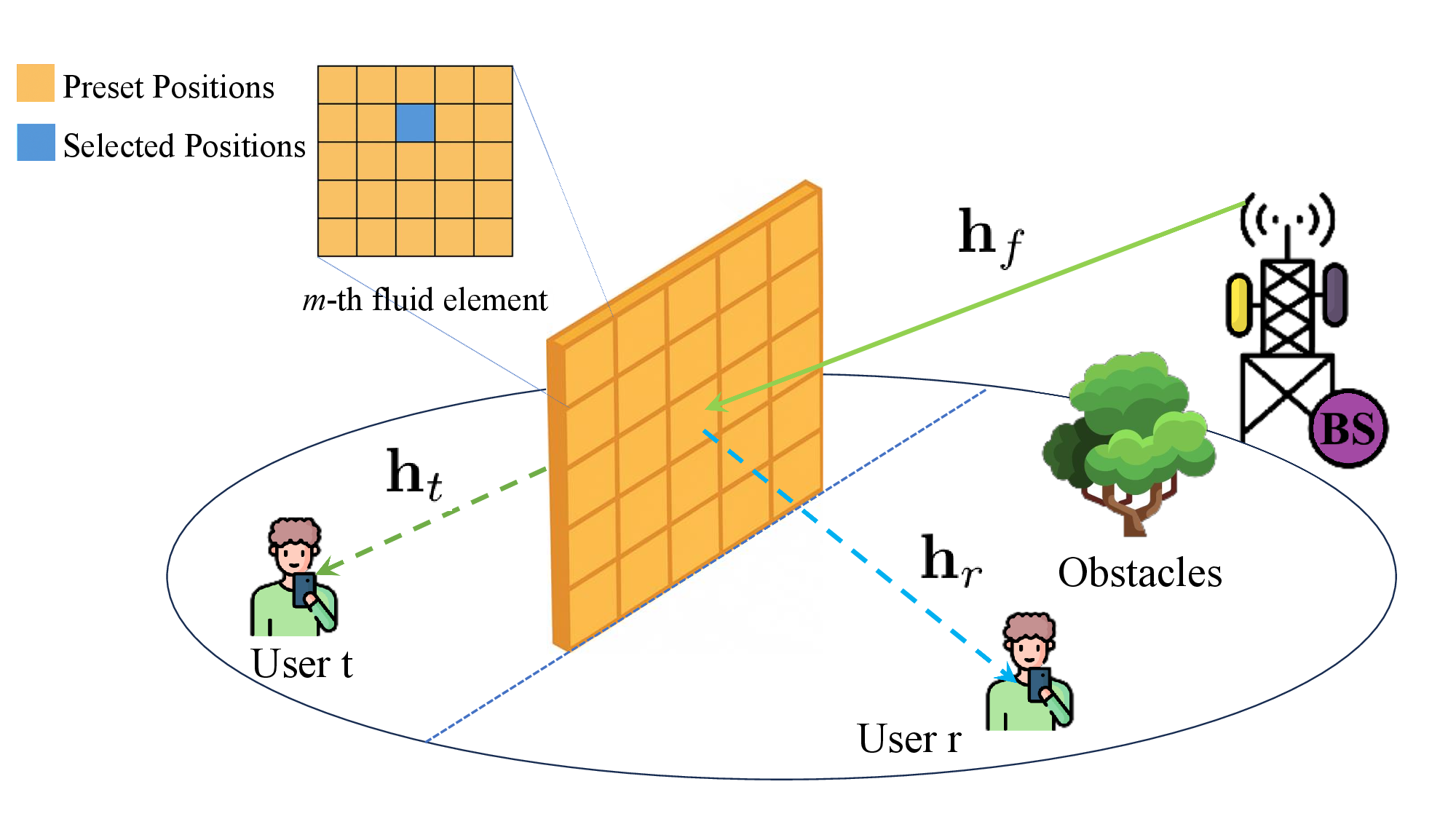}
\caption{The FIRES-aided communication system.}\label{fig_system}
\end{figure}

We consider the downlink wireless communication scenario as shown in Fig.~\ref{fig_system}, where a BS serves two users $u\in\left\{t,r\right\}$ via a FIRES of total aperture $A=A_h\times A_v~\mathrm{m}^2$ and $M$ fluid elements. User $r$ is located in the reflecting half-space of the metasurface, whereas user $t$ is located in the transmitting half-space. We assume that all nodes are equipped with a single fixed-position antenna (FPA) and the direct links between the BS and users are assumed completely blocked. Let the position of the $m$-th fluid element be denoted by $\mathbf{r}_m = \left(x_m, y_m\right)^\mathsf{T}$, and define $\mathcal{S}_m$ as the feasible set of $(x,y)$ coordinates within its designated subarea. For simplicity, we omit the subarea index $m$ and denote the active preset position as $\mathbf r$ instead of $\mathbf r_m$. Each fluid element can switch among $N_m = N_h^m \times N_v^m$ preset positions, where $N_h^m$ and $N_v^m$ represent the numbers of preset positions along the horizontal and vertical directions, respectively, within $\mathcal{S}_m$. Different from a conventional STAR-RIS, whose element locations are fixed, each FIRES element can select its active position $\mathbf r_m$ from the preset set $\mathcal S_m$, while retaining simultaneous transmission/reflection and independent per-side phase control. Hence, a conventional STAR-RIS can be regarded as the special case with a single fixed position for each element. By exploiting its reconfigurability, each fluid element dynamically adjusts its position, power-splitting ratio, and phase shift to shape the propagation environment. Under the ES protocol, the power-splitting coefficient $\beta_{u,m} \in [0,1]$ determines the division of energy between reflection and transmission, i.e., $\beta_{r,m}$ and $\beta_{t,m}$ with $\beta_{r,m}+\beta_{t,m}=1$. For tractability, identical amplitude coefficients are assumed across all elements, i.e., $\beta_{u,m} = \beta_u, \ \forall m$ \cite{ghadi2025fires}. Furthermore, phase adjustments are applied independently for reflection and transmission, denoted by $\phi_m^r$ and $\phi_m^t$, where $\phi_m^u \in (0,2\pi]$. 

{sec:system}
\subsection{Signal and Channel Model}
Let $\mathbf{h}_f \in \mathbb{C}^{M\times 1}$ be the channel between the BS and the FIRES, and let $\mathbf{h}_r \in \mathbb{C}^{M\times 1}$ be the channel between the FIRES and the users. These channels follow a Rician fading model, with a dominant line-of-sight (LoS) path along with scattered non-LoS (NLoS) components in RIS-aided wireless systems \cite{parihar2025perf}. Hence, each hop undergoes Rician fading with path-loss factor $l_q$ and Rician factor $K_q$ for $q \in \left\{f,r,t\right\}$, i.e.,
\begin{align}
\mathbf{h}_q= \sqrt{l_q}\left(\sqrt{\frac{K_q}{K_q+1}}\mathbf{h}_{q,{\rm LoS}} + \sqrt{\frac{1}{K_q+1}}\,\mathbf{h}_{q,{\rm NLoS}}\right),
\end{align}
where $\mathbf{h}_{q,\mathrm{LoS}}$ represents the LoS components, while $\mathbf{h}_{q,\mathrm{NLoS}}$ denotes the NLoS components. The LoS part is characterized through the angles of departure (AoD) and arrival (AoA). For the BS-to-FIRES link, the LoS component at the FIRES is represented by the array response
\begin{align}
\mathbf{h}_{f,{\rm LoS}} \;=\; \mathbf{a}_f\!\left(\psi_{a_f}, \psi_{e_f}, \mathbf{r}\right),
\end{align}
where we have the FIRES receive steering vector
\begin{align}
\left[\mathbf{a}_f\left(\psi_{a_f}, \psi_{e_f}, \mathbf{r}\right)\right]_m=\mathrm{e}^{j\frac{2\pi}{\lambda}\left(x_m\sin \psi_{a_f}\cos \psi_{e_f}+y_m\sin\psi_{e_f}\right)}.
\end{align}
Similarly, the FIRES-to-user LoS channel is written as
\begin{align}
\mathbf{h}_u = \mathbf{a}_u\left(\psi_{a_u}, \psi_{e_u}, \mathbf{r}\right),
\end{align}
where $\psi_{a_u}$ and $\psi_{e_u}$ are the azimuth and elevation AoD toward user $u$, and the corresponding FIRES steering vector is
\begin{align}
\left[\mathbf{a}_u\left(\psi_{a_u}, \psi_{e_u}, \mathbf{r}\right)\right]_m = \mathrm{e}^{j\frac{2\pi}{\lambda}\left(x_m\sin \psi_{a_u}\cos \psi_{e_u}+y_m\sin\psi_{e_u}\right)}.
\end{align}
Given the possible small distance between adjacent preset positions, the spatial correlation between the FIRES elements needs to be considered. For the NLoS components, we model the spatial correlation using Jakes' model. In this regard, we utilize a bijective mapping function $\mathcal{F}: (n_h^m, n_v^m) \mapsto n_m$ for converting the two-dimensional (2D) spatial indices of the metasurface into a 1D linear index. Its inverse, $\mathcal{F}^{-1}(n_m) = (n_h^m, n_v^m)$, allows the retrieval of the original 2D coordinates. Therefore, assuming any two arbitrary preset positions $\tilde{n}_q$ and $\hat{n}_q$, with $\mathcal{F}^{-1}\left(\tilde{n}_q\right)=\left(\tilde{n}_h^q,\tilde{n}_v^q\right)$ and $\mathcal{F}^{-1}\left(\hat{n}_q\right)=\left(\hat{n}_h^q,\hat{n}_v^q\right)$, the spatial correlation matrix takes the form \cite{ghadi2024per}
\begin{multline}\label{eq-rf}
\left[\mathbf{R}_q\right]_{\tilde{n}_q,\hat{n}_q}=\\
\mathrm{sinc}\hspace{-1mm}\left(\frac{2}{\lambda}\sqrt{\left(\frac{\left|\tilde{n}_h^q-\hat{n}_h^q\right|}{L_h-1}A_h\right)^2+\left(\frac{\left|\tilde{n}_v^q-\hat{n}_v^q\right|}{L_v-1}A_v\right)^2}\right),
\end{multline}
where $\mathrm{sinc\left(t\right)=\frac{\sin\left(\pi t\right)}{\pi t}}$. The terms $L_h$ and $L_v$ denote the number of preset positions per row and column of the entire metasurface, respectively.  Hence $L=L_h\times L_v = \sum_{m=1}^M N_m$ is the total number of preset positions. The spatial correlation matrix based on \eqref{eq-rf} can be further written as $\mathbf{R}_q = \mathbf{U}_q \Lambda_q \mathbf{U}_q^H$, 
where $\mathbf{U}_q$ denotes the unitary matrix containing $\mathbf{R}_q$'s eigenvectors and $\mathbf{\Lambda}_q$ is the diagonal matrix containing its eigenvalues. Consequently, an equivalent correlated NLoS draw is $\mathbf{h}_{q,\mathrm{NLoS}} = \mathbf{U}_q\,\mathbf{\Lambda}_q^{1/2}\,\overline{\mathbf{h}}_q$, where $\overline{\mathbf{h}}_q$ represents the uncorrelated small-scale fading component, modeled as independent and identically distributed (i.i.d.)~complex Gaussian variables with zero mean and unit variance. 

As a result, assuming $x$ to be the transmit symbol with $\mathbb{E}[|x|^2]=1$ and total BS power $P$, the received signal of user $u$ for the ES protocol  \cite{mu2022sim} is written as
\begin{align}
y_u = \sqrt{P}\,\mathbf{h}_u^H\mathbf{\Phi}_u\mathbf{h}_fx+z_u, 
\end{align}
where $z_u$ denotes the additive white Gaussian noise (AWGN) with zero mean and variance $\sigma^2$, i.e., $z\sim\mathcal{CN}\left(0,\sigma^2\right)$. The matrix $\mathbf{\Phi}_u\in\mathbb{C}^{M\times M}$ collects the per-side passive coefficients of the FIRES, given by
\begin{align}
\mathbf{\Phi}_{u} \;=\; \mathrm{diag} \big(\sqrt{\beta_u}\,e^{j(\phi_1^u+\epsilon_1^u)},\,\ldots,\,\sqrt{\beta_u}\,e^{j(\phi_M^u+\epsilon_M^u)}\big),
\end{align}
where $\phi_m^u\in[0,2\pi)$ denotes the programmed phase for side $u$, and $\epsilon_m^u$ models the residual phase error, e.g., quantization or hardware inaccuracy. In the absence of errors, coherent combining is achieved by choosing $\phi_m^u=-\theta_m^u$, where $\theta_{m}^u\triangleq\angle\left([\mathbf h_f]_m\right)+\angle\left([\mathbf h_u]_m\right)$ is the per-cell cascaded phase. With imperfect control, the phase coefficient that is actually applied differs from the optimal phase value; hence, this is modeled through an error term as $\phi_m^u+\epsilon_m^u$ so the effective cascaded channel including phase errors becomes
\begin{align}
h^{\mathrm{eff}}_u \triangleq \mathbf h_u^H\mathbf\Phi_u\mathbf h_f=\sqrt{\beta_u}\sum_{m=1}^M \left|\left[\mathbf h_u\right]_m\right|\left|\left[\mathbf h_f\right]_m\right| \mathrm{e}^{j\epsilon_m^u}.\label{eq:coh-sum}
\end{align}  
We define $a_m\triangleq\left|\left[\mathbf h_u\right]_m\right|\left|\left[\mathbf h_f\right]_m\right|\ge 0$ and the phase-aligned cascaded gain as $H_u(\mathbf r)\;\triangleq\;\sum_{m=1}^M a_m$, which depends on the FIRES positions $\mathbf r$ through the array responses. If $\{\epsilon_m^u\}$ are i.i.d., zero-mean, and independent of $\{a_m\}$, the normalized phasor sum concentrates around its mean. We therefore model the loss by the deterministic attenuation defined as
\begin{align}
\chi_u \,\triangleq\, \Bigg|\frac{1}{M}\sum_{m=1}^M e^{j\epsilon_m^u}\Bigg|\,\approx\, \big|\mathbb E[e^{j\epsilon_m^u}]\big| \,\in (0,1], \label{eq:chi_def}
\end{align}
so that $|\mathbf h_u^H\mathbf\Phi_u\mathbf h_f|\approx \sqrt{\beta_u}\,\chi_u\,H_u(\mathbf r)$. 

Invoking $\chi_u=\big|\mathbb{E}[e^{j\epsilon_m^u}]\big|$, the attenuation admits closed-forms for Gaussian jitter and $Q$-level uniform quantization error models, yielding
\begin{align}
	\chi_u \;=\;
	\begin{cases}
		e^{-\sigma_\phi^2/2}, 
		& \text{ } \epsilon_m^u \sim \mathcal N(0,\sigma_\phi^2),\\[2mm]
		\displaystyle \bigg|\frac{\sin(\pi/Q)}{\pi/Q}\bigg|, 
		& \text{ } \epsilon_m^u \sim \mathcal U\!\big[-\tfrac{\pi}{Q},\,\tfrac{\pi}{Q}\big].
	\end{cases}
	\label{eq:chi_models}
\end{align}
Both expressions satisfy $\chi_u \to 1$ as $\sigma_\phi^2 \to 0$ or $Q \to \infty$, corresponding to ideal phase control. Thus, the instantaneous signal-to-noise ratio (SNR) at user $u$ is approximated by
\begin{align}
\gamma_u \,\approx\, \frac{P}{\sigma^2}\,\beta_u\left(\chi_u\,H_u(\mathbf r)\right)^2. \label{eq:SNRu}
\end{align}
It should be noted that expression \eqref{eq:SNRu} is exact for $\chi_u=1$ and provides a tight deterministic equivalent for i.i.d.~errors as $M$ increases. Specifically, it separates three roles: resource allocation is represented by $P$ and $\beta_u$, hardware and control quality is represented by $\chi_u$, and propagation and geometry are represented by $H_u$, which depends on $\mathbf r$. 

\vspace{-2mm}
\section{Coverage Characterization and Analysis}\label{sec:coverage}
Here, we derive the far-field coverage radii\footnote{Throughout this paper, we consider the far-field and that all distances are larger than the Rayleigh distance $D_{\rm Ray}=2A/\lambda. $} that expose key scalings, and then specialize them to OMA and NOMA with SNR/signal-to-interference-plus-noise ratio (SINR) targets.

\vspace{-2mm}
\subsection{Closed-Form Far-Field LoS Coverage Bound}\label{subsec:closed-form}
We denote $d_f$ as the BS-to-FIRES distance and $D_u$ as the FIRES-to-user distance along the dominant LoS. For far-field LoS, the $m$-th channel magnitude for the corresponding links can be approximated as $\left|[\mathbf{h}_f]_m\right| \approx \sqrt{\rho_0}\,d_f^{-\alpha/2}$ and  $\left|[\mathbf{h}_u]_m\right| \approx \sqrt{\rho_0}\,D_u^{-\alpha/2}$ for all $m$, where $\rho_0$ is the unit-distance reference gain \cite{Rappaport2015}. Hence, the phase-aligned cascaded gain becomes
\begin{align}
H_u(\mathbf{r}) \approx M\,\rho_0\,d_f^{-\alpha/2}D_u^{-\alpha/2}.
\end{align}
Then the SNR in \eqref{eq:SNRu} reduces to
\begin{align}
\gamma_u\approx \frac{P}{\sigma^2}\,\beta_u\,\chi_u^2\,M^2\,\rho_0^2\,d_f^{-\alpha}D_u^{-\alpha}.\label{eq:SNR_farfield}
\end{align}

\begin{theorem}[FIRES LoS Coverage Radius]\label{thm:LoS}
Under far-field LoS and ES protocol, any user $u\in\{r,t\}$ is covered if
\begin{align}
D_u \;\le\; \Bigg(\frac{P}{\sigma^2}\;\frac{\beta_u\,\chi_u^2\,M^2\,\rho_0^2}{\gamma_{\mathrm{th},u}\,d_f^{\alpha}}\Bigg)^{\!1/\alpha},\label{eq:Du_bound}
\end{align}
where $\gamma_{\mathrm{th},u}$ is the SNR threshold. 
\end{theorem}

\begin{IEEEproof}
Direct substitution of the far-field $H_u$ into \eqref{eq:SNRu} and solving $\gamma_u\ge\gamma_{\mathrm{th},u}$ gives \eqref{eq:Du_bound}.
\end{IEEEproof}

The same template applies to OMA and NOMA by inserting the scheme-specific resource shares, e.g., time in OMA via $\tau_u$, power in NOMA via $p_u$, and using the corresponding SNR and SINR targets $\gamma_{\mathrm{th},u}^\zeta$, where $\zeta\in\left\{\mathrm{O,N}\right\}$ indicates if OMA and NOMA schemes are considered. 

\begin{remark}
Theorem \ref{thm:LoS} represents a tractable far-field LoS specialization of the general position-dependent channel model introduced in Section \ref{sec:system}. Under this approximation, all FIRES elements experience approximately identical propagation magnitudes, and the explicit dependence on the element positions vanishes. Hence, the result serves as an analytical benchmark and reveals the scaling law
	\[
	D_u \propto
	P^{1/\alpha}
	M^{2/\alpha}
	\beta_u^{1/\alpha}
	\chi_u^{2/\alpha}
	\left(\gamma_{\mathrm{th},u}\right)^{-1/\alpha}
	d_f^{-1}. 
	\]
This shows, for example, that increasing the number of active elements provides an $M^{2/\alpha}$ coverage gain, while phase control impairments inevitably reduce the achievable coverage through $\chi_u^{2/\alpha}$. The complete  channel model is retained in the optimization developed in Sections \ref{sec:max_coverage} and \ref{sec:pos_opt}.
\end{remark}

\begin{corollary}[OMA coverage bound]\label{cor:OMA}
Consider OMA with time fraction $\tau_u\in(0,1]$ allocated to user $u$ and  target rate $R_u^{\rm tar}$ bps/Hz. Under the considered far-field LoS model, OMA user $u$ is covered whenever
\begin{align}
D_u^\mathrm{O} \;\le\;
\left(\frac{P\,\chi_u^2\,M^2\,\rho_0^2}{\sigma^2\gamma_{\mathrm{th},u}^{\rm O}\,d_f^{\alpha}}\right)^{1/\alpha},\label{eq:Du_OMA}
\end{align}
where the required SINR threshold is $\gamma_{\mathrm{th},u}^{\rm O}=2^{\,R_u^{\rm tar}/\tau_u}-1$.
\end{corollary}

\begin{IEEEproof}
In OMA, $R_u=\tau_u\log_2(1+\gamma_u)\ge R_u^{\rm tar}$ which is equivalent to $\gamma_u\ge \gamma_{\mathrm{th},u}^{\rm O}$. Considering the far-field SNR from \eqref{eq:SNR_farfield} and rearranging $\gamma_u\ge \gamma_{\mathrm{th},u}^{\rm O}$ for $D_u$ gives the result.
\end{IEEEproof}

\begin{corollary}[NOMA coverage bounds]\label{cor:NOMA}
Consider two-user power-domain NOMA with BS power fraction $p_u\in(0,1)$ and $p_r+p_t=1$. Under the considered far-field LoS model, the following  
sufficient conditions ensure coverage for the candidate decoding
order in which user \(r\) performs SIC and user \(t\) is decoded first,
respectively,
\begin{align}
D_{r}^\mathrm{N} \;\le\;
\left(\frac{Pp_r\beta_r\,\chi_r^2\,M^2\,\rho_0^2}{\sigma^2d_f^{\alpha}\gamma_{\mathrm{th},r}^\mathrm{N}}\right)^{1/\alpha}\label{eq:Du_NOMA_strong}
\end{align}
and 
\begin{align}
D_t^\mathrm{N} \;\le\;
\left(\frac{P\beta_t\,\chi_t^2\,M^2\,\rho_0^2\left(p_t-p_r\gamma_{\mathrm{th},t}^\mathrm{N}\right)}{\sigma^2d_f^{\alpha}\gamma_{\mathrm{th},t}^\mathrm{N}}\;\right)^{1/\alpha},\label{eq:Du_NOMA_weak}
\end{align}
where $\gamma_{\mathrm{th},u}^\mathrm{NOMA}$ is the SINR threshold of NOMA user $u$.
\end{corollary}

\begin{IEEEproof}
For the candidate decoding order considered here, let user \(r\) be the SIC capable user and user \(t\) be decoded first. Under \eqref{eq:SNR_farfield}, we define $S_u \triangleq \frac{P}{\sigma^2}\,\beta_u\,\chi_u^2\,M^2\,\rho_0^2\,d_f^{-\alpha}\,D_u^{-\alpha}$, and therefore, the SINR of the strong user after perfect SIC is $\gamma_r=p_rS_r$. Enforcing $\gamma_r\ge \gamma_{\mathrm{th},r}^\mathrm{N}$ and solving for $D_r^\mathrm{N}$ yields \eqref{eq:Du_NOMA_strong}. For the weak user, the SINR is given by $\gamma_t=p_tS_t/(p_rS_t+1)$. The inequality $\gamma_t\ge \gamma_{\mathrm{th},t}^\mathrm{N}$ is equivalent to $(p_t-p_r\gamma_{\mathrm{th},t})S_t\ge \gamma_{\mathrm{th},t}^\mathrm{N}$, which requires $p_t>p_r\gamma_{\mathrm{th},t}^\mathrm{N}$ and leads to \eqref{eq:Du_NOMA_weak}. The corresponding bounds for the reverse decoding order follow directly
	by interchanging the roles of users \(r\) and \(t\).
\end{IEEEproof}

\begin{remark}
Setting $\chi_u=1$ represents ideal phase control. In OMA, it is customary to set $\beta_u=1$ during user-$u$'s slot, although \eqref{eq:Du_OMA} holds for any fixed $\beta_u$. In NOMA, \eqref{eq:Du_NOMA_strong} and \eqref{eq:Du_NOMA_weak} reduce to Theorem~\ref{thm:LoS} when all BS power is allocated to a single user (or when only one user is scheduled).
\end{remark}

\vspace{-2mm}
\section{Coverage Maximization}\label{sec:max_coverage}
In this section, we first formulate the OMA coverage maximization problem and illustrate that for fixed preset positions, the inner problem reduces to a 1D search over the time fraction variable. This structure is later exploited within the outer PSO-based layout optimization. 
Then, we formulate the NOMA coverage design, consider both possible decoding orders, state the corresponding SIC feasibility conditions, and reveal that for fixed positions, the inner search becomes a low-dimensional numerical optimization over the power and ES variables.

\vspace{-2mm}
\subsection{OMA Coverage Maximization }\label{subsec:OMA}
In the OMA protocol, users are served in orthogonal time slots using the full transmit power $P$ and full FIRES energy allocation, i.e., $\beta_t=\beta_r=1$, during their respective slots. Therefore, consider the OMA scheme with time\footnote{The adopted OMA-time division multiple access (TDMA) formulation can also be considered as the time-splitting (TS) operation of FIRES for the two-user topology. In particular, during the fraction $\tau$, the surface is configured to serve the reflecting-side user with full energy allocation, i.e., $\beta_r=1$, whereas during the remaining fraction $1-\tau$, it serves the transmitting-side user with $\beta_t=1$. In contrast, under ES, the reflecting and transmitting functions operate simultaneously subject to $\beta_r+\beta_t=1$. Hence, TS avoids the ES loss by dedicating the full surface response to one side at a time, at the expense of a time-allocation penalty, whereas ES enables simultaneous service of both half-spaces by dividing the incident energy between them.} fraction $\tau$ allocated to user $r$ and $\left(1-\tau\right)$ to user $t$, so that the OMA rates are, respectively, defined as
\begin{align}
R_r^{\rm OMA}&=\tau\,\log_2\!\left(1+\frac{P}{\sigma^2}\,\left(\chi_r\,H_r(\mathbf r)\right)^2\right),\\
R_t^{\rm OMA}&=(1-\tau)\,\log_2\!\left(1+\frac{P}{\sigma^2}\,\left(\chi_t\,H_t(\mathbf r)\right)^2\right).
\end{align}
Given the target rate $R_u^{\rm tar}$, the quality-of-service (QoS) constraints are, respectively, denoted as
\begin{align}
R_r^{\rm OMA}&\ge R_r^{\rm tar},\\
R_t^{\rm OMA}&\ge R_t^{\rm tar}.\label{eq:OMA_qos_phase}
\end{align}

Now, let $D_r$ and $D_t$ be the coverage radii on the reflecting and transmitting sides, respectively. We define $D_{\rm tot}\triangleq D_r+D_t$ as the aggregate bilateral coverage extent across the two half-spaces. 
The total coverage design for the OMA scheme is formulated as
\begin{subequations}\label{prob:OMA_phase}
\begin{align}
\max_{\;D_r,D_t,\,\tau,\,\mathbf r} &~D_\mathrm{tot}^\mathrm{O} \\
\text{s.t.} &~ \tau\,\log_2\!\Big(1+\tfrac{P}{\sigma^2}\,\chi_r^2\,H_r(\mathbf r)^2\Big)\;\ge\; R_r^{\rm tar}, \label{eq:OMA_c1_phase}\\
&~ (1-\tau)\,\log_2\!\Big(1+\tfrac{P}{\sigma^2}\,\chi_t^2\,H_t(\mathbf r)^2\Big)\;\ge\; R_t^{\rm tar}, \label{eq:OMA_c2_phase}\\
&~ \tau\in(0,1), \\
&~ \mathbf r_m\in\mathcal S_m, \|\mathbf r_m-\mathbf r_{m'}\|_2\ge D, \forall m\neq m',
\end{align}
where $D$ is the minimum spacing between two fluid elements.
\end{subequations}

\subsubsection{Inner structure for fixed positions}
For fixed $\mathbf r$, we define $S_u=\frac{P}{\sigma^2}\,\chi_u^2\,H_u(\mathbf r)^2>0$. Hence, the QoS constraints become $\tau\,\log_2\!\big(1+S_r\big)\ge R_r^{\rm tar}$ and $(1-\tau)\,\log_2\!\big(1+S_t\big)\ge R_t^{\rm tar}$. For fixed $\mathbf r$, the QoS constraints impose lower and upper bounds on the admissible resource fraction $\tau$. Hence, the corresponding QoS-induced bounds are, respectively, given by
\begin{align}
\tau_{\rm LB}&= \frac{R_r^{\rm tar}}{\log_2\!\big(1+S_r\big)},\\
\tau_{\rm UB}&
=
1-
\frac{R_t^{\rm tar}}
{\log_2\!\left(1+S_t\right)}, \label{eq:tau_star_phase}
\end{align}
in which feasibility requires
$
	0<\tau_{\rm LB}\leq\tau_{\rm UB}<1$. 

\begin{theorem}[1D split for OMA with phase errors]\label{thm:oma_1d_phase}
For fixed $\mathbf r$, the OMA resource-allocation problem reduces to a 1D search over $\tau\in(0,1)$. The feasible values of $\tau$ are determined by the two QoS constraints, in which the constraint of user $r$ is monotonically increasing in $\tau$, while that of user $t$ is monotonically decreasing in $\tau$.
\end{theorem}

\begin{IEEEproof}
The QoS constraints in OMA can be written as $\tau\log_2\!\big(1+S_r(\mathbf r)\big)\ge R_r^{\rm tar}$ and $(1-\tau)\log_2\!\big(1+S_t(\mathbf r)\big)\ge R_t^{\rm tar}$, with  $S_u=\frac{P}{\sigma^2}\,\chi_u^2\,H_u(\mathbf r)^2$. Since $S_r(\mathbf r)$ and $S_t(\mathbf r)$ are fixed for a given FIRES configuration, the first constraint imposes the lower bound $\tau \ge\frac{R_r^{\rm tar}}{\log_2\!\big(1+S_r(\mathbf r)\big)}$, whereas the second constraint imposes the upper bound $\tau \le 1-\frac{R_t^{\rm tar}}{\log_2\!\big(1+S_t(\mathbf r)\big)}$. Therefore, the feasible interval is $\tau_{\rm LB}	\le \tau \le 	\tau_{\rm UB}$. Hence, for fixed $\mathbf r$, the OMA inner problem only requires a 1D search over this feasible interval. This completes the proof.
\end{IEEEproof}

\vspace{-2mm}
\subsection{NOMA Coverage Optimization}\label{subsec:NOMA_rstrong}
We consider power-domain NOMA with superposition $x=\sqrt{p_r}\,s_r+\sqrt{p_t}\,s_t$, where $s_r,s_t$ have unit power, $p_u\ge 0$, and $p_r+p_t\le 1$. We also define $\widetilde G_u(\mathbf r)\triangleq \frac{\big(\chi_u\,H_u(\mathbf r)\big)^2}{\sigma^2}$ which embeds propagation, FIRES geometry, and phase-error loss.  
Under the ES protocol with \(\beta_r+\beta_t=1\), we first consider the candidate decoding order in which user \(r\) performs SIC. The corresponding SINRs are
\begin{align}
\gamma_t^{\rm N} &= \frac{\beta_t\,p_t\,\widetilde G_t}{\beta_t\,p_r\,\widetilde G_t+1},\label{eq:noma_t_weak}\\
\gamma_{r\to t}^{\rm N} &=\frac{\beta_r p_t \widetilde G_r}{\beta_r p_r \widetilde G_r+1},\label{eq:noma_r_decode_t}\\
\gamma_r^{\rm N} &= \beta_r\,p_r\,\widetilde G_r,\label{eq:noma_r_strong}
\end{align}
with the achievable rates given by
\begin{align}
R_t^{\rm N}&=\log_2\!\left(1+\frac{\beta_t\,p_t\,\widetilde G_t}{\beta_t\,p_r\,\widetilde G_t+1}\right),\\
R_r^{\rm N}&=\log_2\!\left(1+\beta_r\,p_r\,\widetilde G_r\right).
\end{align}
Given the target rate $R_u^{\rm tar}$, the QoS and SIC constraints are defined as
\begin{subequations}\label{eq:noma_constraints_phase_rstrong}

\begin{align}
R_t^{\rm tar}&\leq\log_2\!\left(1+\frac{\beta_t p_t \widetilde G_t}{\beta_t p_r \widetilde G_t+1}\right),\\
R_r^{\rm tar}&\leq\log_2\!\left(1+\beta_r p_r \widetilde G_r\right),\\
R_t^{\rm tar}&\leq\log_2\!\left(1+\frac{\beta_r p_t \widetilde G_r}{\beta_r p_r \widetilde G_r+1}\right).
\end{align}
\end{subequations}
Therefore, the total coverage region problem for the NOMA scheme is formulated as
\begin{subequations}\label{prob:NOMA_phase_rstrong}
\begin{align}
\max_{\;D_r,D_t,\,\beta_r,\beta_t,\atop p_r,p_t,\,\mathbf r} &~D_\mathrm{tot}^\mathrm{N} \\
\text{s.t.} &~ \text{\eqref{eq:noma_constraints_phase_rstrong}},\\
&~p_r+p_t\le 1,~\mbox{for }p_r,p_t\ge 0,\\
&~\beta_r+\beta_t=1,~\beta_u\in[0,1], \\
&~\mathbf r_m\in\mathcal S_m, \|\mathbf r_m-\mathbf r_{m'}\|_2\ge D, \forall m\neq m'.
\end{align}
\end{subequations}

\subsubsection{Decoding order and feasibility}
For the adopted decoding order, where user $r$ performs SIC, a feasible NOMA allocation must satisfy the following conditions:
\begin{align}
\frac{\beta_r\,p_t\,\widetilde G_r}{\beta_r\,p_r\,\widetilde G_r+1}&\ge \gamma_{\mathrm{th},t}^\mathrm{N},\\
\frac{\beta_t\,p_t\,\widetilde G_t}{\beta_t\,p_r\,\widetilde G_t+1}&\ge \gamma_{\mathrm{th},t}^\mathrm{N},\label{eq:SIC_pair_rstrong}
\end{align}
together with $\beta_r\,p_r\,\widetilde G_r\ge \gamma_{\mathrm{th},r}^\mathrm{N}$, where $\gamma_{\mathrm{th},u}^\mathrm{N}=2^{R_u^\mathrm{tar}}-1$. Note that the first inequality in \eqref{eq:SIC_pair_rstrong} enforces that, as seen at the strong user $r$, the desired layer for $t$ exceeds the residual interference from $r$ by the SIC margin $\gamma_{\mathrm{th},t}^\mathrm{N}$. 

\begin{remark}[Feasibility conditions]
For fixed $\mathbf r$, assume that user $r$ is selected as the SIC-capable user. A resource allocation $(p_r,p_t,\beta_r,\beta_t)$ satisfying $p_r,p_t\ge0$, $p_r+p_t\le1$, $\beta_r+\beta_t=1$, and $\beta_u\in[0,1]$ is feasible for the adopted decoding order if the QoS and SIC conditions in \eqref{eq:SIC_pair_rstrong}, together with $\beta_r p_r \widetilde G_r\ge\gamma_{\rm th,r}^{\rm N}$, are satisfied.
\end{remark}

The reverse candidate decoding order, in which user \(t\) performs SIC, is obtained by interchanging the roles of \(r\) and \(t\) in \eqref{eq:noma_t_weak}--\eqref{eq:SIC_pair_rstrong}. For every candidate FIRES configuration, both decoding
	orders are evaluated subject to their respective QoS and SIC feasibility
	conditions. If both orders are feasible, the one yielding the larger
	total coverage is retained; if only one is feasible, that order is used.

\subsubsection{Inner optimization for fixed positions}
	For fixed \(\mathbf r\), the effective channel gains
	\(\widetilde G_u(\mathbf r)\) are constants. For each of the two
	candidate decoding orders, the inner problem involves only the resource
	allocation variables \((\beta_r,p_r)\), with
	\(\beta_t=1-\beta_r\) and \(p_t=1-p_r\), subject to the corresponding
	QoS and SIC feasibility conditions. Since both problems are
	low-dimensional, they can be solved efficiently for each candidate
	FIRES configuration, and the feasible decoding order yielding the
	larger total coverage is retained.

\vspace{-2mm}
\section{Position-Aware Optimization: Bi-Level PSO}\label{sec:pos_opt}
In this section, we maximize the FIRES coverage by searching over discrete element positions under spacing and one-active-preset constraints. An outer PSO\footnote{The choice of PSO is motivated by the structure of the outer FIRES position optimization. The active element positions are selected from discrete preset sets, while the resulting objective is nonconvex and generally nonsmooth due to preset projection, spacing constraints, feasibility penalties, and the embedded OMA or NOMA resource allocation. Consequently, gradient information with respect to the position variables is not readily available. PSO provides a convenient derivative-free framework for this setting, since continuous particle coordinates can be directly projected onto the admissible presets and constraint violations can be incorporated through repair and penalty mechanisms. Moreover, particle evaluations are mutually independent and can therefore be parallelized at the BS or edge controller. PSO is not the only applicable metaheuristic, but it offers a simple and effective match to the mixed discrete and continuous structure considered here.} proposes positions $\mathbf r\triangleq[\mathbf r_1,\ldots,\mathbf r_M]$ (one active preset per subarea), while an inner OMA/NOMA split certifies QoS feasibility and returns the objective $D_\mathrm{tot}(\mathbf r)$. The coupling between geometry and performance enters through the phase-aligned cascaded gains $H_u(\mathbf r)$ and the active-set NLoS covariance $\mathbf R_q^{(\mathrm{act})}(\mathbf r)$.

\vspace{-2mm}
\subsection{Bi-Level Structure}
We adopt a bi-level scheme. The outer loop searches over $\mathbf r$; the inner loop solves a low-dimensional resource allocation problem in the protocol variables given $\mathbf r$.
\begin{itemize}
\item \textbf{Outer objective.} We aim to maximize the total coverage $D_\mathrm{tot}(\mathbf r)=D_r(\mathbf r)+D_t(\mathbf r)$, where $D_u(\mathbf r)$ is evaluated from the position-dependent effective gain $H_u(\mathbf r)$, while the far-field expressions in Section~\ref{sec:coverage} are used as analytical benchmarks and scaling references.
\item \textbf{Inner problems.} The inner problems employ $\gamma_u(\mathbf r)=\frac{P}{\sigma^2}\,\beta_u\,\big(\chi_u\,H_u(\mathbf r)\big)^2$, consistent with Section~\ref{subsec:closed-form}.
\end{itemize}
 
\subsubsection{OMA} For fixed $\mathbf r$, define $S_u(\mathbf r)\!=\!\frac{P}{\sigma^2}\chi_u^2 H_u(\mathbf r)^2$. The QoS constraints are $\tau\log_2(1+S_r)\!\ge\!R_r^{\rm tar}$ and $(1-\tau)\log_2(1+S_t)\!\ge\!R_t^{\rm tar}$. As shown in Theorem \ref{thm:oma_1d_phase}, the inner problem reduces to a 1D search over $\tau\!\in\!(0,1)$ with $\tau_{\rm LB}$ and $\tau_{\rm UB}*$.

\subsubsection{NOMA} With BS power fractions \(p_u\geq0\) and
\(p_r+p_t\leq1\), and ES split \(\beta_r+\beta_t=1\), both
candidate decoding orders are evaluated for fixed \(\mathbf r\).
For each order, the inner problem is a low-dimensional search in
\((\beta_r,p_r)\), with \(p_t=1-p_r\) and
\(\beta_t=1-\beta_r\), subject to the corresponding QoS and SIC
feasibility conditions. The feasible order yielding the larger
total coverage is retained.

\vspace{-2mm}
\subsection{Channel Update for a Given \texorpdfstring{$\mathbf r$}{r}}\label{sec:vb}
For a candidate FIRES configuration
	\(\mathbf r=[\mathbf r_1,\ldots,\mathbf r_M]\), only the
	\(M\) selected active positions are required for channel generation.
	Accordingly, for hop \(q\in\{f,r,t\}\), we directly construct the
	active spatial covariance matrix
	\(\mathbf R_q^{(\mathrm{act})}(\mathbf r)\in\mathbb C^{M\times M}\)
	from the pairwise distances between the selected positions as
	\begin{align}
		\left[\mathbf R_q^{(\mathrm{act})}(\mathbf r)\right]_{m,m'}
		=
		\mathrm{sinc}\!\left(
		\frac{2}{\lambda}
		\left\|\mathbf r_m-\mathbf r_{m'}\right\|_2
		\right),
		\label{eq:active_covariance}
	\end{align}
	where $	\quad m,m'=1,\ldots,M$. This is equivalent to selecting the entries associated with the
	active presets from the full grid Jakes correlation model in \label{eq-rf},
	but avoids explicitly constructing or storing the full
	\(L\times L\) covariance matrix. Then, a correlated NLoS realization is  generated from
	\(\mathbf R_q^{(\mathrm{act})}(\mathbf r)\), while the LoS steering
	vectors are recomputed using the selected positions \(\mathbf r\).
	The resulting Rician channel is subsequently used to evaluate
	\(H_u(\mathbf r)\) and hence \(S_u(\mathbf r)\).

\vspace{-2mm}
\subsection{Constraint Handling}
The feasible set is given as
\begin{align}
\mathbf r_m\in\mathcal S_m, \|\mathbf r_m-\mathbf r_{m'}\|_2\ge D~(\forall m\neq m'),
\end{align}
with exactly one active preset per subarea. We enforce these via a projection $\mathcal P(\cdot)$ onto $\{\mathcal S_m\}$ and a penalty
\begin{equation}
\mathfrak B_{\rm space}(\mathbf r)=\sum_{m<m'}\mathbf 1\big[\|\mathbf r_m-\mathbf r_{m'}\|_2<D\big].
\end{equation}
The penalized outer objective is $\mathcal J(\mathbf r)=\mathcal G(\mathbf r)-\mu\,\mathfrak B_{\rm space}(\mathbf r)$, where $\mathcal G$ is $D_\mathrm{tot}$,  $\mu\!>\!0$ is the penalty weight, and where $\mathfrak B_{\rm space}$ penalizes spacing violations.

\subsection{PSO-Based Outer Solver}\label{subsec:pso}
We maximize the penalized objective
\begin{equation}
\mathcal J(\mathbf r)=D_\mathrm{tot}(\mathbf r)-\mu_{\rm space}\,\mathfrak B_{\rm space}(\mathbf r)-\mu_{\rm q}\,\mathfrak B_{\rm q}(\mathbf r),
\end{equation}
where $\mathfrak B_{\rm q}$ penalizes QoS/SIC violations.

\subsubsection{Encoding and projection}
The FIRES aperture is partitioned into $M$ disjoint subareas $\{\mathcal S_m\}$, each with a discrete preset set $\mathcal P_m=\{\mathbf p_{m,1},\ldots,\mathbf p_{m,L_m}\}$. A particle encodes a continuous surrogate $\mathbf y_m\in[0,1]^2$ per subarea. The geometric projection $\Pi_{\rm geom}$ maps $\mathbf y_m$ to a physical point in $\mathcal S_m$; the discrete projection $\Pi_{\rm disc}$ snaps it to the nearest preset:
\begin{equation}
\Pi(\mathbf y_m)\triangleq\arg\min_{\mathbf p\in\mathcal P_m}\|\Pi_{\rm geom}(\mathbf y_m)-\mathbf p\|_2.
\end{equation}
Collecting the active choices gives $\mathbf r=\big[\Pi(\mathbf y_1),\ldots,\Pi(\mathbf y_M)\big]$ with one active preset per subarea.

\subsubsection{Constraint handling}
The minimum spacing $D$ is enforced by a repair-and-penalize mechanism. After snapping, if any $\|\mathbf r_m-\mathbf r_{m'}\|_2<D$, we greedily reassign one of the two to its next-best preset in $\mathcal P_m$; remaining conflicts contribute to
\begin{equation}
\mathfrak B_{\rm space}(\mathbf r)=\sum_{m<m'}\mathbf 1\!\left[\|\mathbf r_m-\mathbf r_{m'}\|_2<D\right].
\end{equation}
If the inner problem (OMA or NOMA) is infeasible at $\mathbf r$, e.g., QoS or SIC violated, we set
\begin{equation}
\mathfrak B_{\rm q}(\mathbf r)=1+\sum_{u\in\{r,t\}}\max\big\{0,R_u^{\rm tar}-R_u(\mathbf r)\big\};
\end{equation}
otherwise $\mathfrak B_{\rm q}(\mathbf r)=0$. The penalties $\mu_{\rm space}$ and $\mu_{\rm q}>0$ are chosen large enough to dominate $\mathcal F$ when violated; we use an adaptive schedule increasing $\mu$ if violations persist.

\subsubsection{Particle updates}
Let $\mathbf y[p]\in[0,1]^{2M}$ be particle $p$'s position surrogate and {\bf v}[p] its velocity. At iteration $t$,
\begin{multline}
\mathbf v^{(t+1)}[p]=w^{(t)}\,\mathbf v^{(t)}[p]+c_1 r_1\big(\mathbf y_{\rm pbest}[p]-\mathbf y^{(t)}[p]\big)\\
+c_2 r_2\big(\mathbf y_{\rm gbest}-\mathbf y^{(t)}[p]\big)
\end{multline}
and
\begin{align}
\mathbf y^{(t+1)}[p]&=\mathrm{clip}\big(\mathbf y^{(t)}[p]+\mathbf v^{(t+1)}[p],0,1\big)
\end{align}
with $r_1,r_2\sim\mathcal U(0,1)$ i.i.d. We use a decreasing inertia $w^{(t)}=w_{\min}+(w_{\max}-w_{\min})\frac{T-t}{T}$, with $w_{\max}\approx 0.9,\;w_{\min}\approx 0.3$, and velocity clamping $\|\mathbf v^{(t)}[p]\|_\infty\le v_{\max}$, e.g., $v_{\max}=0.2$. Positions are snapped via $\Pi$ before evaluation.

\subsubsection{Initialization and seeding}
Half of the swarm is seeded heuristically by selecting, in each subarea, the preset maximizing a local surrogate of
\begin{align}
\max_{\mathbf p\in\mathcal P_m}\sum_{u\in\{r,t\}}\big|\mathbf a_u(\mathbf p)^H\big|\,\big|\mathbf a_f(\mathbf p)\big|,
\end{align}
and the remainder is randomized uniformly. This accelerates convergence and improves diversity.

\subsubsection{Scoring a particle}
Given $\mathbf r$, we update channels and compute $H_u(\mathbf r)$:
\begin{itemize}
\item \emph{LoS terms:} array responses at the snapped positions.
\item NLoS terms: construct the active
	\(M\times M\) covariance
	\(\mathbf R_q^{(\mathrm{act})}(\mathbf r)\)
	directly from the selected element positions using
	\eqref{eq:active_covariance}, then draw
	\(\mathbf h_{q,\mathrm{NLoS}}\sim
	\mathcal{CN}(\mathbf 0,\mathbf R_q^{(\mathrm{act})})\)
	and form the Rician channel \(\mathbf h_q\).

\end{itemize}
Optionally, average $D_\mathrm{tot}(\mathbf r)$ over $N_\mathrm{MC}$ Monte Carlo fading draws (small $N_\mathrm{MC}$, e.g., $N_\mathrm{MC}=5$) to reduce noise. The inner OMA or NOMA program is then solved with SNRs $\gamma_u(\mathbf r)$. The returned objective $D_\mathrm{tot}(\mathbf r)$ and penalties yield $\mathcal J(\mathbf r)$.

\subsubsection{Stopping and complexity}
We stop at $T$ iterations or if the global best stalls for $T_{\rm stall}$ iterations. Per iteration, the cost is $\mathcal O\!\big[N_p\,[M\log L_{\max}+C_{\rm inner}]\big]$. With $N_\mathrm{MC}$ Monte-Carlo averages per particle, the per-iteration cost becomes $\mathcal{O}\!\big[N_p\,N_\mathrm{MC}\,[M\log L_{\max}+C_{\rm inner}]\big]$; equivalently, runtime is approximately $N_\mathrm{MC}$ times higher than in the single-draw case. All runs use fixed random seeds for reproducibility.

We implement the outer search using PSO in Algorithm \ref{alg:pso_pos}.\footnote{In practical IoT deployments, the proposed optimization is executed centrally at the BS or an associated edge controller rather than at the IoT devices. Moreover, position reconfiguration need not be performed at every channel coherence interval. The FIRES element positions can be updated on a slower timescale according to user geometry and long-term or averaged channel conditions, while the inner resource allocation variables may be adapted more frequently. This separation substantially reduces the control and computational burden on energy-constrained IoT devices, which only require conventional channel estimation or feedback support.}

In the adopted discrete preset architecture, position reconfiguration is realized by switching the active location within each subarea. Therefore, the associated hardware overhead is mainly determined by the switching and control circuitry, the number of available presets, and the reconfiguration frequency. The energy consumption is implementation dependent and not assigned a fixed value in the present system-level model.

\begin{algorithm}[!t]
\caption{Outer PSO for Position-Aware FIRES Design}\label{alg:pso_pos}
\begin{small}
\begin{algorithmic}[1]
\STATE \textbf{Input:} subareas $\{\mathcal S_m\}$, spacing $D$, swarm size $N_p$, iterations $T$, inertia $w$, gains $c_1,c_2$.
\STATE Initialize particles $\{\mathbf r^{(0)}[p]\}_{p=1}^{N_p}$ with one active preset per subarea; set velocities $\{\mathbf v^{(0)}[p]\}$.
\FOR{$t=0$ to $T-1$}
\FOR{$p=1$ to $N_p$}
\STATE Project to feasibility: $\mathbf r^{(t)}[p]\leftarrow \mathcal P(\mathbf r^{(t)}[p])$.
\STATE Update channels using $\mathbf R_q^{(\mathrm{act})}(\mathbf r^{(t)}[p])$ and compute $H_u(\mathbf r^{(t)}[p])$.
\STATE \textbf{Inner solve:}  
OMA 1D split (Thm.~\ref{thm:oma_1d_phase}), or NOMA low-dimensional search under both candidate decoding orders; retain the feasible order yielding the larger total coverage.
\STATE Evaluate $\mathcal D_\mathrm{tot}(\mathbf r^{(t)}[p])$ and $\mathcal J(\mathbf r^{(t)}[p])$.
\STATE Update personal and global bests.
\ENDFOR
\STATE PSO velocity/position updates:
\begin{align}
\mathbf v^{(t+1)}[p]&=w\,\mathbf v^{(t)}[p]+c_1 r_1(\mathbf r_{\rm pbest}[p]-\mathbf r^{(t)}[p])\notag\\
&+c_2 r_2(\mathbf r_{\rm gbest}-\mathbf r^{(t)}[p]),\notag\\
\mathbf r^{(t+1)}[p]&=\mathcal P\big(\mathbf r^{(t)}[p]+\mathbf v^{(t+1)}[p]\big).\notag
\end{align}
\ENDFOR
\STATE \textbf{Output:} $\mathbf r^*=\mathbf r_{\rm gbest}$ together with the corresponding optimal resource allocation, i.e., $\tau^*$ for OMA or $(\beta_r^*,p_r^*,p_t^*)$ for NOMA, and the achieved total coverage $D_{\rm tot}(\mathbf r^*)$.
\end{algorithmic}
\end{small}
\end{algorithm}

\subsection{Complexity and Convergence}\label{subsec:complexity}
We now characterize the precomputation cost, per-iteration PSO cost, inner-loop cost for OMA/NOMA, and overall complexity, and summarize convergence behavior.

 \subsubsection{Active covariance construction}
	For each candidate configuration, the active covariance matrix
	\(\mathbf R_q^{(\mathrm{act})}(\mathbf r)\in\mathbb C^{M\times M}\)
	is constructed directly from the pairwise distances between the
	selected element positions. Its construction requires
	\(O(M^2)\) operations and \(O(M^2)\) memory per hop.
	A factorization of the active covariance matrix for correlated
	channel generation requires \(O(M^3)\) operations.

\subsubsection{Per-iteration cost}
Each PSO iteration evaluates \(N_p\) particles. For each particle,
	constructing the active covariance matrix requires \(O(M^2)\)
	operations, while its factorization requires \(O(M^3)\).
	Once the factorization is available, each correlated NLoS channel
	realization requires \(O(M^2)\) operations. Hence, with
	\(N_{\rm MC}\) Monte Carlo channel realizations per particle,
	the channel generation cost scales as
	\(O(M^3+N_{\rm MC}M^2)\).
	The computation of the phase-aligned gains \(H_u(\mathbf r)\)
	requires an additional \(O(M)\) operations.

\subsubsection{Inner problems} 
OMA reduces to a 1D search over $\tau$ with $	\tau_{\rm LB}$ in closed form; a bracketing/golden-section search with $N_{\rm eva}$ evaluations yields $\mathcal O(N_{\rm eva})$ per particle, i.e., each evaluation is constant-time given $S_u$. On the other hand, NOMA is solved by a low-dimensional search over $(\beta_r,p_r)$ with $p_t=1-p_r$. Since only two scalar variables are involved, the computational cost of the NOMA inner search is negligible compared with the channel update and PSO evaluation.

\subsubsection{Overall complexity} 
	Accounting for the active covariance construction, its factorization,
	the Monte Carlo channel generation, and the inner resource allocation,
	the per-iteration complexity is
	\begin{align}
		O\!\left(
		N_p\left[
		M^3+N_{\rm MC}M^2+C_{\rm inner}
		\right]
		\right),
		\label{eq:overall_complexity}
	\end{align}
	where \(C_{\rm inner}=O(N_{\rm eva})\) for OMA and remains
	small for the low-dimensional NOMA search. Over \(T\) iterations,
	the overall complexity becomes
	\begin{align}
		O\!\left(
		T N_p\left[
		M^3+N_{\rm MC}M^2+C_{\rm inner}
		\right]
		\right).
	\end{align}
	The covariance-related memory requirement is \(O(M^2)\) per hop,
	since only the active covariance matrix is constructed and stored.

\subsubsection{Convergence} The outer PSO utilizes inertia decay and velocity clamping, ensuring stable trajectories and monotonic improvement of the incumbent global best. As a metaheuristic for a nonconvex discrete-continuous landscape, PSO does not ensure global optimality, but empirically converges reliably under the adopted seeding and penalties. Phase-control imperfections enter only through $\chi_u$ inside $S_u(\mathbf r)$, hence are fully accounted for in the objective evaluations used by PSO.

Representative runtimes and convergence statistics are summarized in Table \ref{tab:complexity}. The wall-clock values in Table~\ref{tab:complexity} correspond to the bound-based
	PSO objective and characterize the runtime of the outer position
	search and inner resource-allocation procedure. Therefore, they  do
	not include the covariance factorization and Monte Carlo channel
	generation used for the correlated Rician
	evaluation. Also, it should be noted that the scalability of the proposed framework is primarily determined by the number of FIRES elements, the PSO population size, the iteration budget, and the dimension of the inner resource allocation problem. Increasing $M$ enlarges the outer search dimension and the cost of each candidate evaluation, although the fitness evaluations of different particles are independent and can be parallelized. For a multiuser extension, additional cascaded channels, QoS constraints, and resource-allocation variables would further increase the per-particle computational cost. Accordingly, larger surface sizes and a greater number of users are expected to increase the overall computational burden and may require a larger search effort for convergence.

\begin{table*}[t]
\centering
\caption{PSO runtime summary for FIRES coverage optimization (bound-based objective).}\label{tab:complexity}
\vspace{-0.5cm}
\begin{adjustbox}{center,margin=1em}
\begin{tabular}{|c|c|c|c|c|c|c|c|c|}
\hline
Mode & $M$ & $N_p$ & $T$ & $N_h^{\rm sub}\!=\!N_v^{\rm sub}$ & TotTime [s] & Time/iter [ms] & It$_{99}$ & $D_\mathrm{tot}^{\rm best}$ [m] \\
\hline\hline
OMA  & $36$ & $30$ & $80$ & $100$ & $0.20595$ & $2.5743$ & $22$ & $2.0654$ \\
NOMA & $36$ & $30$ & $80$ & $100$ & $0.18478$ & $2.3097$ & $2$ & $2.8247$ \\
OMA  & $64$ & $30$ & $100$ & $100$ & $0.85380$ & $8.5380$ & $24$ & $2.9850$ \\
NOMA & $64$ & $30$ & $100$ & $100$ & $0.80513$ & $8.0513$ & $3$ & $4.1150$ \\
OMA  & $36$ & $60$ & $100$ &  $60$ & $0.60131$ & $6.0131$ & $12$ & $2.0654$ \\
NOMA & $36$ & $60$ & $100$ & $60$ & $0.49604$ & $4.9604$ & $29$ & $2.8790$ \\
\hline
\end{tabular}
\end{adjustbox}
\end{table*}

\vspace{-2mm}
\section{Numerical Evaluation}\label{sec:sim}
In this section, we present numerical results for FIRES with position-aware optimization under both OMA and NOMA. We consider conventional STAR-RIS as the benchmark because it provides an architecturally equivalent full-space reference for FIRES. Both architectures support simultaneous transmission and reflection, whereas STAR-RIS lacks element-wise position reconfigurability. For fair comparison, the STAR-RIS benchmark uses the same number of elements, aperture, transmit power, and resource allocation rules as FIRES. Therefore, the resulting performance difference directly reflects the benefit by the spatial reconfigurability of FIRES. Unless stated otherwise, all parameters follow Table \ref{tab:parameter}. FIRES activates one preset per subarea; the outer positions are selected via PSO, while  
the inner step optimizes the OMA time fraction or the NOMA
	ES/power allocation while evaluating both possible SIC decoding orders.

\begin{table}[t]
\centering
\caption{Simulation Parameters.}\label{tab:parameter}
\vspace{-0.5cm}
\begin{adjustbox}{center,margin=1em}
\begin{tabular}{|c||c|}
\hline
Parameter & Value\\
\hline
$f_c$ & $3.5$ GHz\\
$\alpha$ & $2.1$ \\
$\sigma^2$ & $-114$ dBm\\
$d_f$ & $50$ m\\
$\rho_0$ & $-13.3$ dB \\
$P$ & $30$ dBm \\	
$R^\mathrm{tar}_u$ & $1$ bps/Hz \\
$A$ & $1$ $\rm m^2$ \\
$M$ & $\left\{16,36,64\right\}$ \\
$w$ & $0.4$ \\
$c_1$ & $0.5$ \\
$c_2$ & $0.5$ \\
$N_h^m$ & $100$ \\
$N_v^m$ & $100$ \\
$D$ & $\lambda/2$ \\
$K_q$ & $5$ \\
$T$ & $60$ \\
$N_p$ & $30$ \\
\hline
\end{tabular}
\end{adjustbox}
\end{table}

\begin{figure}[!t]
\centering
\includegraphics[width=0.9\columnwidth]{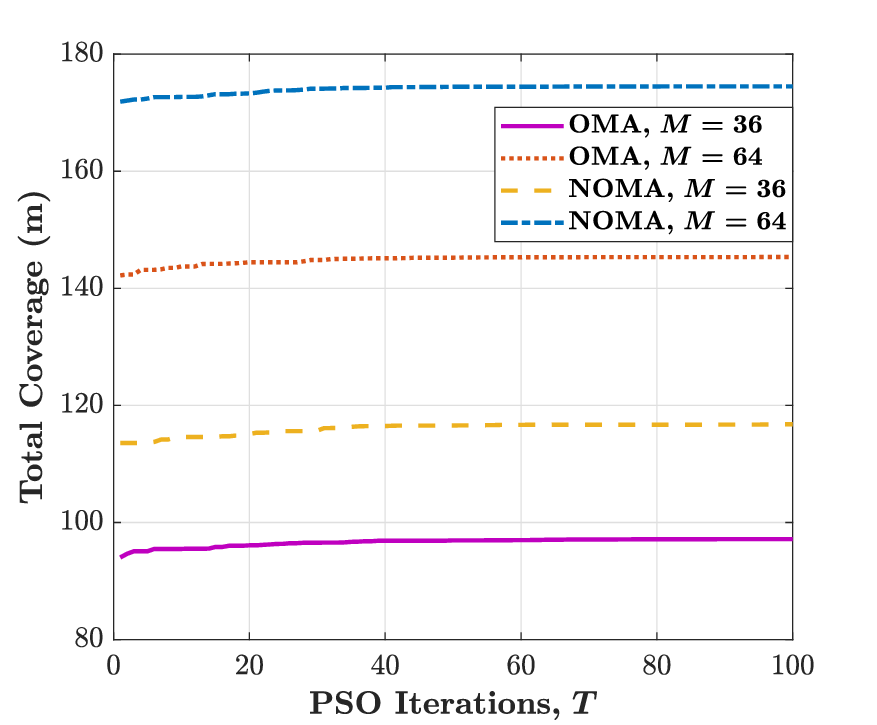}
\caption{Total coverage $D_\mathrm{tot}$ versus PSO iterations $T$. }\label{fig:it}
\end{figure}

Fig.~\ref{fig:it} shows the total coverage versus PSO iterations $T$ for $M\!=\!36$ and $M\!=\!64$. All curves rise quickly during the first $20-40$ iterations and then flatten, indicating near-convergence by $T\approx 60$. Increasing the number of active elements shifts the curves upward while leaving the convergence rate essentially unchanged. The residual gains after $T\approx 60$ are marginal, so $T\in[60,100]$ is a practical budget for the outer loop.

\begin{figure*}[!t]
\centering
\includegraphics[width=2\columnwidth]{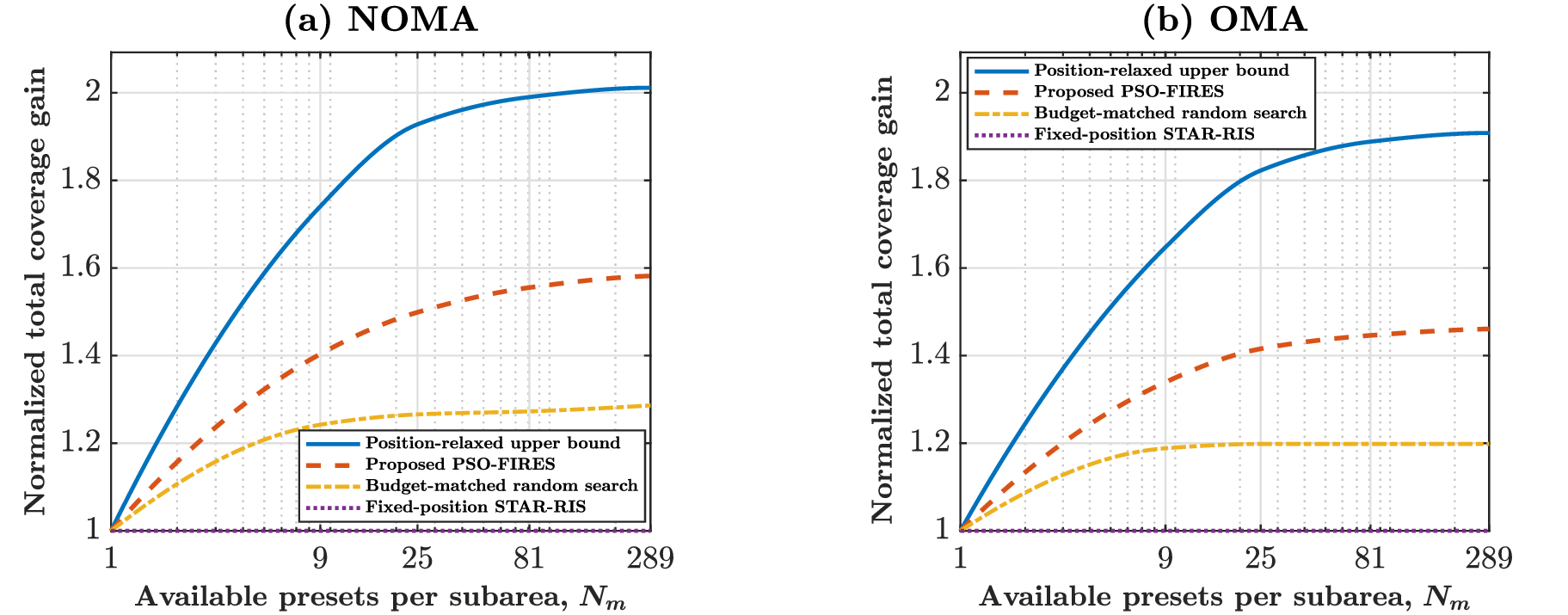}
\caption{Normalized total coverage versus the number of available presets per subarea for different position-selection benchmarks under (a) NOMA and (b) OMA, with $M=36$, $N_p=30$, and $T=60$. }\label{fig:bench}
\end{figure*}

To further assess the effectiveness of the proposed position optimization, Fig.~\ref{fig:bench} compares PSO-based FIRES with two additional references as the number of available presets per subarea increases. For each access scheme, the results are normalized by the corresponding fixed-position STAR-RIS baseline to highlight the relative gain provided by spatial reconfigurability. The equal budget random search uses the same number of objective function evaluations as PSO, while the decoupled position upper bound independently maximizes the two side-specific spatial gains without enforcing a common realizable element position. For both NOMA and OMA, the proposed PSO-based FIRES consistently outperforms the equal-budget random search and the fixed-position STAR-RIS, while remaining below the relaxed upper bound. This shows that the coverage improvement stems not only from providing additional spatial DoF, but also from effectively exploiting them through position-aware optimization.

To further identify the physical origin of this position reconfiguration gain, we performed a controlled ablation study for \(M=36\) and \(N_m=289\). Under the correlated Rician model, the optimized FIRES provides relative coverage gains of approximately \(57.7\%\) and \(52.5\%\) over the corresponding fixed position STAR-RIS for NOMA and OMA, respectively. When the spatial correlation of the NLoS component is removed while retaining position-dependent NLoS variations, the corresponding gains remain \(54.6\%\) and \(54.5\%\). Hence, spatial correlation reduction has only a secondary effect on the observed gain. In contrast, when the NLoS component is removed and only the LoS
	component is retained, the position reconfiguration gain vanishes. This confirms that the dominant benefit of FIRES arises from exploiting position-dependent variations of the scattered/NLoS channel. Finally, phase control errors primarily introduce the common attenuation factor \(\chi_u\). For example, for Gaussian phase jitter with \(\sigma_\phi^2=0.5~\mathrm{rad}^2\), the coverage scaling factor \(\chi_u^{2/\alpha}\) is approximately \(0.788\), corresponding to about a \(21.2\%\) reduction in absolute coverage. Since the same attenuation model applies to FIRES and STAR-RIS, robustness to phase control errors
	is not the primary source of the relative FIRES gain.

\begin{figure}[!t]
\centering
\includegraphics[width=0.9\columnwidth]{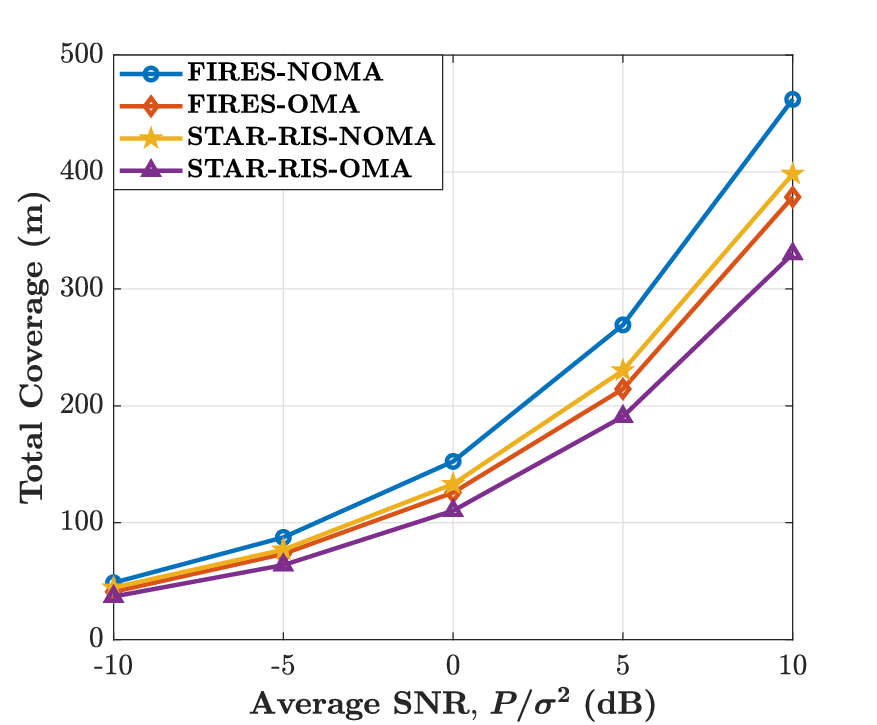}
\caption{Total coverage $D_\mathrm{tot}$ versus the average SNR $P/\sigma^2$.}\label{fig:snr}
\end{figure}

In Fig.~\ref{fig:snr}, we present the total coverage performance in terms of the average SNR, $P/\sigma^2$. We see that $D_\mathrm{tot}$ grows as the average SNR increases because the radius obeys $D_u \propto (P/\sigma^2)^{1/\alpha}$, so a higher SNR expands the feasible region nonlinearly.  
 We also see that FIRES traces sit above STAR-RIS since selecting one active preset per subarea allows FIRES to exploit position-dependent channel variations and improve the phase-aligned cascaded gain, thereby increasing the effective array factor even with the same \(M\). The gap widens at high SNR where geometry limits performance. Also, for a fixed surface, NOMA exceeds OMA more clearly as SNR increases because the SIC margin grows and the BS can use a more asymmetric power split to serve both users in the full time slot, whereas OMA keeps a pre-log penalty from time/frequency partitioning. At low SNR, the schemes are noise-limited, power splits are constrained, and the gap between NOMA and OMA is modest, but the separation becomes pronounced at moderate-to-high SNR.

\begin{figure}[!t]
\centering
\includegraphics[width=0.9\columnwidth]{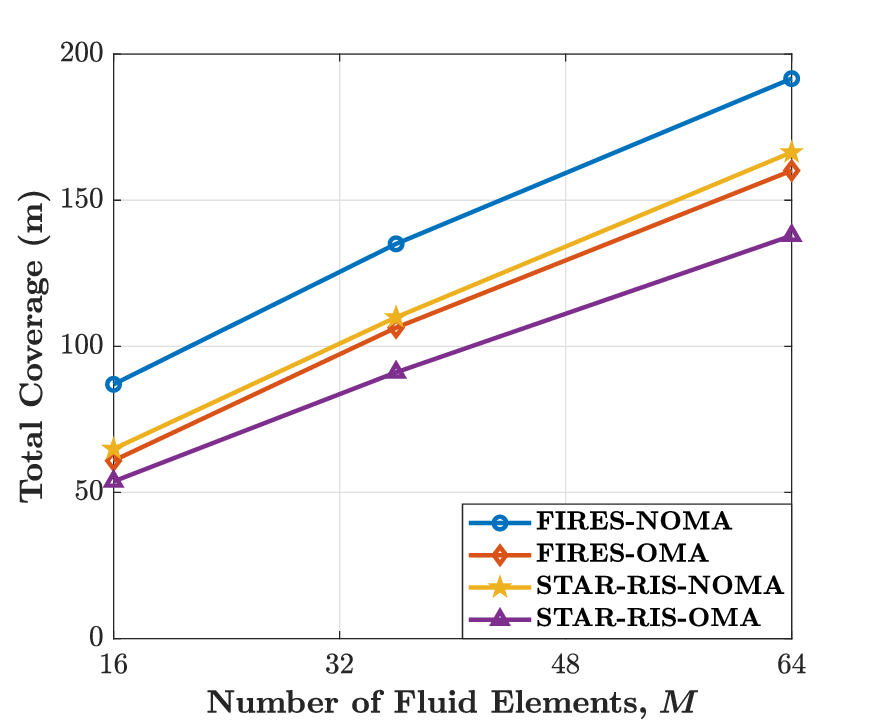}
\caption{Total coverage $D_\mathrm{tot}$ versus the number of fluid elements $M$.}\label{fig:m}
\end{figure}

Fig.~\ref{fig:m} illustrates the impact of the number of fluid elements $M$ on the total coverage performance. We see that coverage improves monotonically with increasing $M$. This behavior arises because the cascaded gain scales with the effective aperture, while the SNR grows proportionally to $M^{2}$, leading to $D_u \propto M^{2/\alpha}$ under the far-field model.  
Moreover, FIRES consistently outperforms its STAR-RIS counterparts because position agility enables the surface to exploit favorable position-dependent variations of the cascaded channel. Spatial correlation reduction provides an additional benefit, but the ablation results indicate that it is not the dominant source of the observed gain. Additionally, the performance gap widens slightly for larger $M$, as the optimizer exploits the increased DoF available for active preset placement. At this operating point, the NOMA curves, as expected, remain above the OMA curves for both the FIRES and STAR-RIS cases owing to the superposition strategy.

\begin{figure}[!t]
\centering
\includegraphics[width=0.9\columnwidth]{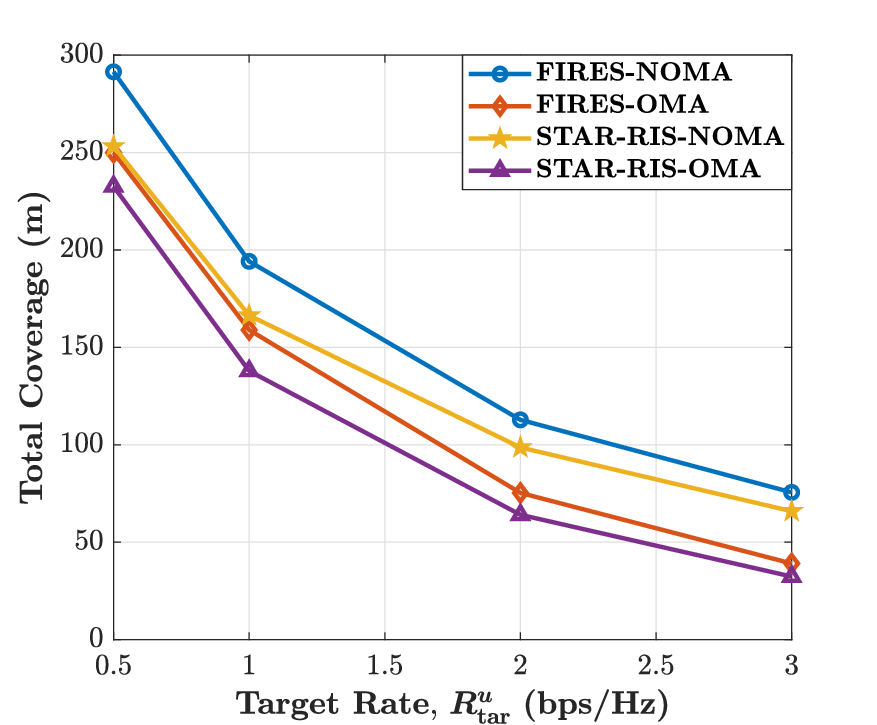}
\caption{Total coverage $D_\mathrm{tot}$ versus the target rate $R^\mathrm{tar}_u$.}\label{fig:qos}
\end{figure}

Fig.~\ref{fig:qos} illustrates how the target rate $R_u^{\rm tar}$ affects the total coverage $D_{\rm tot}=D_r+D_t$. We observe that the total coverage decreases monotonically as the target rate increases since the required SINR $\gamma_{\mathrm{th},u}$ grows exponentially with $R^\mathrm{tar}_u$ and the coverage bound scales as $D_u\propto \gamma_{\mathrm{th},u}^{-1/\alpha}$. For a given surface, NOMA outperforms OMA at low to moderate rates, since superposition with SIC exploits the broadcast gain and avoids time/frequency partitioning, yielding more effective power on both links. As $R_u^{\rm tar}$ becomes high, the NOMA advantage narrows because the weak user's SIC condition $p_t>p_r\gamma_{{\rm th},t}$ becomes stringent, driving power splits toward OMA-like allocations, and the curves converge at high target rates.

\begin{figure}[!t]
\centering
\includegraphics[width=0.9\columnwidth]{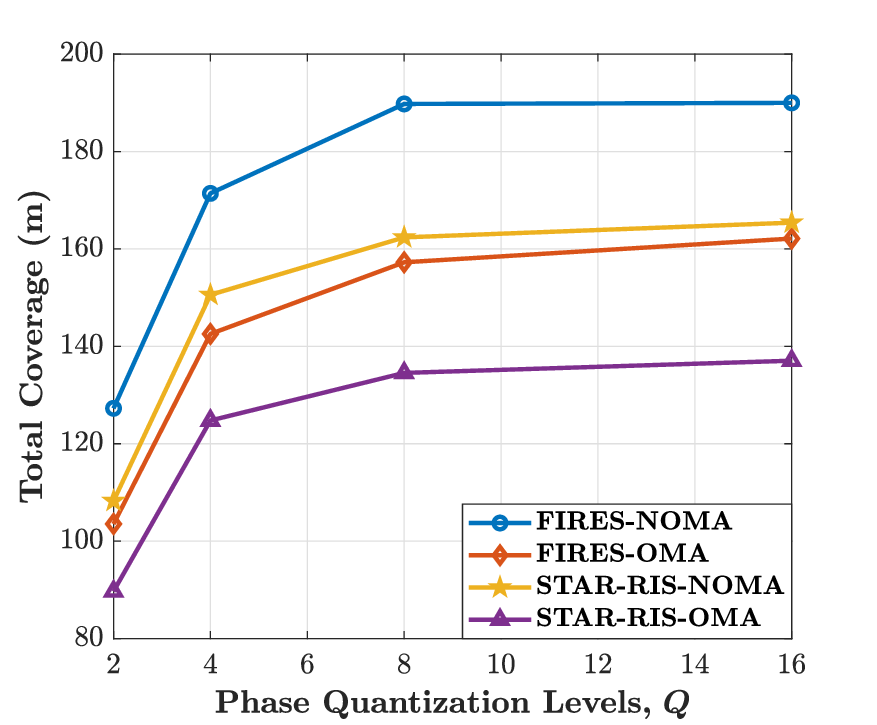}
\caption{Total coverage $D_\mathrm{tot}$ versus the phase quantization level $Q$.}\label{fig:quan}
\end{figure}

Fig.~\ref{fig:quan} shows the total coverage versus the number of phase-quantization levels $Q$. Coverage increases monotonically with $Q$ for all schemes and then saturates, mainly because finer quantization lets the metasurface elements align phases more accurately, so the reflected and transmitted wavefronts add nearly in co-phase at the users. This behavior follows from the phase-error attenuation in \eqref{eq:chi_models} approaching one as $Q$ grows and from the bound $D_u \propto \chi^{2/\alpha}$. The largest gain occurs when moving from $Q=2$ to $Q=4$, where $\chi$ increases sharply, whereas the gain from $Q=8$ to $Q=16$ is incremental, indicating that about three to four bits of phase control are already sufficient. The ordering is consistent across $Q$, with FIRES-NOMA achieving the largest coverage; the advantage comes from FIRES's position flexibility, which improves coherent combining toward both sides.

\begin{figure}[!t]
\centering
\includegraphics[width=0.9\columnwidth]{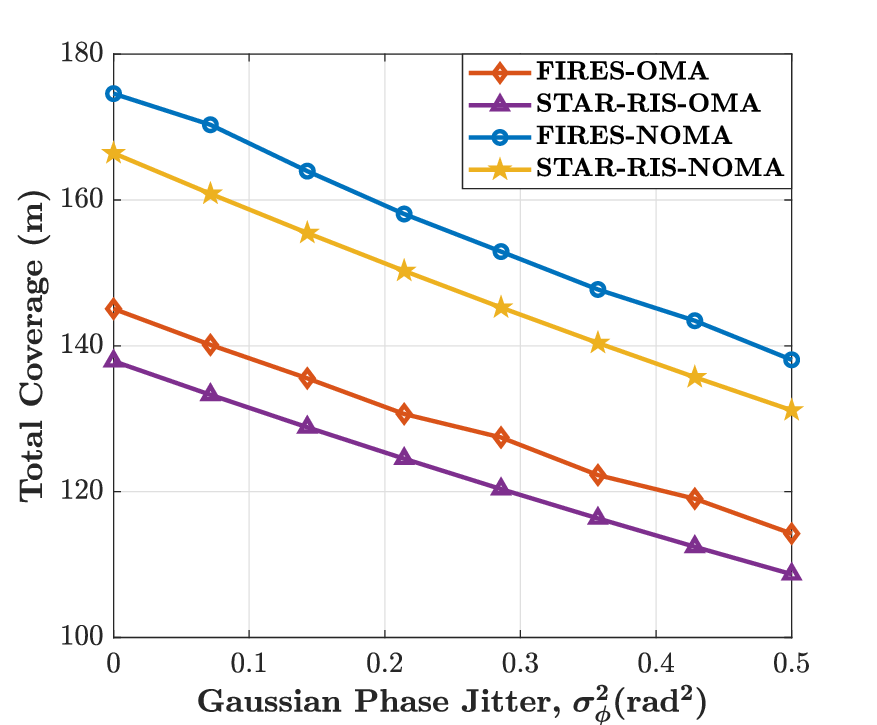}
\caption{Total coverage $D_\mathrm{tot}$ versus the Gaussian phase jitter $\sigma^2_{\phi}$.}\label{fig:jit}
\end{figure}

In Fig.~\ref{fig:jit}, we investigate the impact of Gaussian phase jitter with variance $\sigma_\phi^2$ on the total coverage. Coverage decreases as $\sigma_\phi^2$ grows because random phase errors reduce coherent combining; under the standard small-jitter model the effective phasor scales as $\chi=\exp(-\sigma_\phi^2/2)$, so the bound yields $D_\mathrm{tot}\propto \chi^{2/\alpha}\approx \exp(-\sigma_\phi^2/\alpha)$. At $\sigma_\phi^2=0.5\ \mathrm{rad}^2$, FIRES-NOMA drops from about $175$ m to around $140$ m, a $20\%$ reduction, while FIRES-OMA and the two STAR-RIS baselines show similar losses. The ordering remains unchanged across the sweep; FIRES configurations stay above STAR-RIS, as expected, because they retain stronger coherent gain even under phase jitter, and the NOMA pairs keep a modest edge since both links benefit from superposition with SIC.

\begin{figure}[!t]
\centering
\includegraphics[width=0.9\columnwidth]{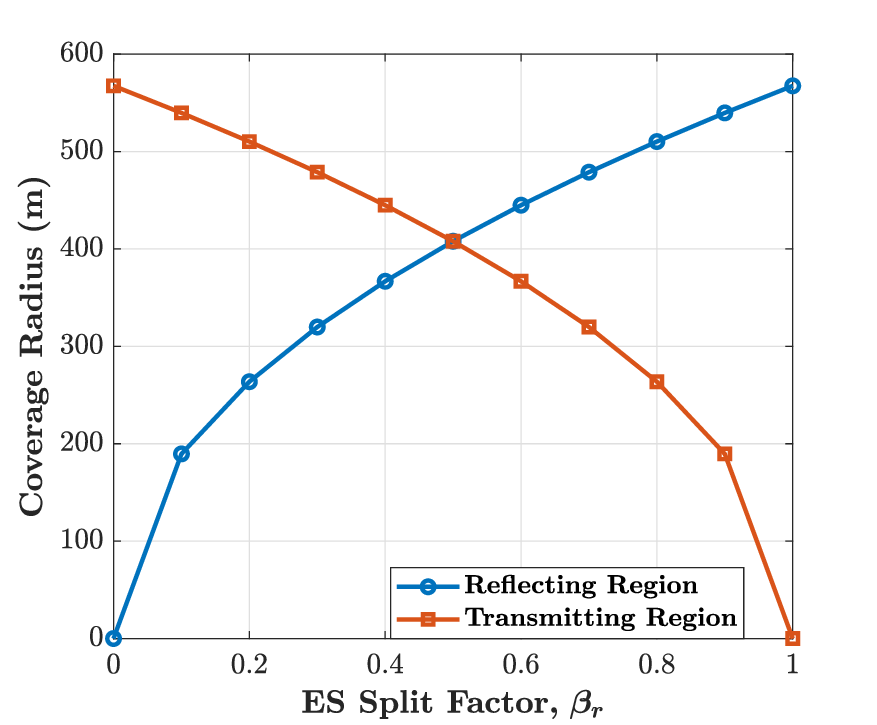}
\caption{Total coverage $D_\mathrm{tot}$ versus the ES split factor $\beta_r$.}\label{fig:beta}
\end{figure}

Fig.~\ref{fig:beta} depicts the total coverage on the transmitting and reflecting sides as a function of the ES split factor $\beta_r$ with $\beta_t=1-\beta_r$. As $\beta_r$ increases, $D_r$ grows while $D_t$ decreases, producing the expected power-allocation tradeoff. The curves are smooth and sublinear because the far-field bound yields $D_r \propto \beta_r^{1/\alpha}$ and $D_t \propto (1-\beta_r)^{1/\alpha}$ with path-loss exponent $\alpha>2$. Under the symmetric setup used here both radii intersect near $\beta_r\approx 0.5$; with unequal targets or channel factors the intersection would shift toward the side requiring more power. 

\begin{figure}[!t]
\centering
\includegraphics[width=0.9\columnwidth]{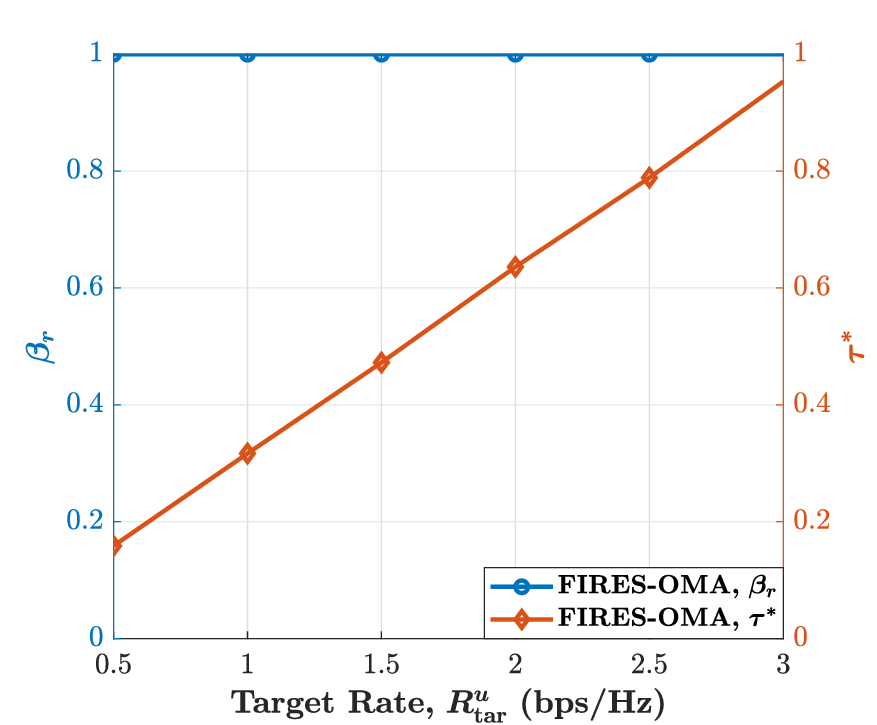}
\caption{OMA energy-allocation coefficient $\beta_r=1$ and optimal time fraction $\tau_{\rm LB}$ versus target rate $R_u^{\rm tar}$.}\label{fig:rate}
\end{figure}

Fig.~\ref{fig:rate} illustrates the optimal OMA time-fraction allocation $\tau_{\rm LB}$ as the target rate $R_u^{\mathrm{tar}}$ varies. Since in OMA, the full FIRES aperture and transmit power are devoted to one user per slot, the ES split factor is fixed as $\beta_r = 1$. The time fraction $\tau_{\rm LB}$ increases with the target rate, reflecting that higher rate demands require a longer slot for the stronger link to meet $R_r^{\mathrm{tar}}$. The nearly linear trend confirms that time allocation is the primary control parameter in OMA once ES is inactive.

\begin{figure}[!t]
\centering
\includegraphics[width=0.9\columnwidth]{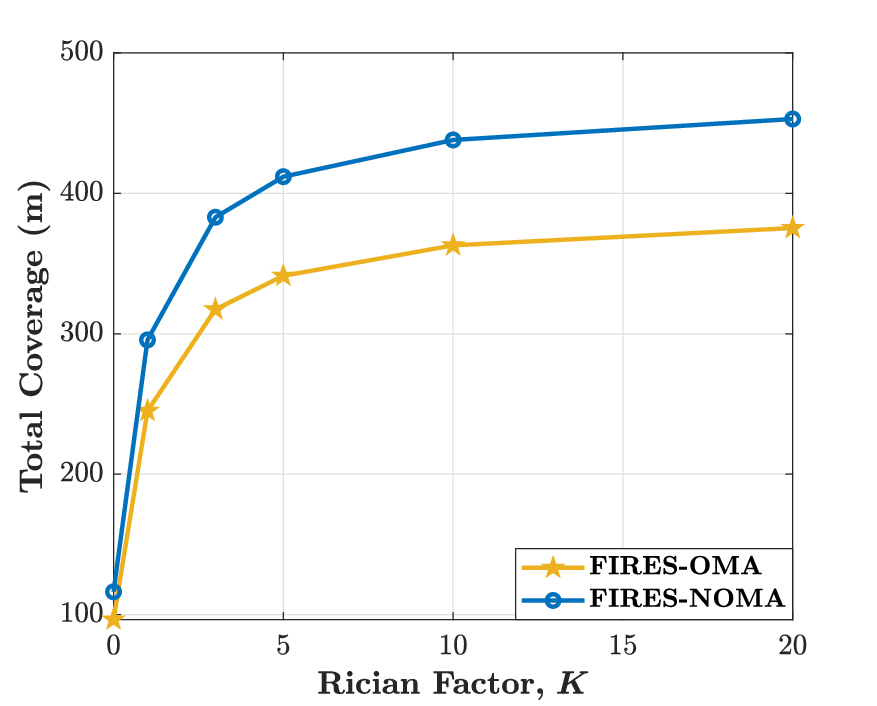}
\caption{Total coverage $D_\mathrm{tot}$ versus the Rician factor $K$.}\label{fig:ric}
\end{figure}

Fig.~\ref{fig:ric} shows total coverage versus the Rician factor $K$ for FIRES with OMA and NOMA. Coverage increases with $K$ and then flattens. As the specular component strengthens, coherent combining over the surface becomes more effective, so the cascaded gain grows roughly with the LoS fraction $K/(K+1)$. The curves exhibit early saturation because once the channel is dominated by LoS, further increases in $K$ bring only small additional coherent gain. Across the sweep, FIRES-NOMA remains above FIRES-OMA since superposition with SIC avoids time partitioning and lets both users benefit from the specular path simultaneously. The gap is more visible at low-to-moderate $K$ where interference management matters; it narrows as $K$ becomes large and both schemes are limited mainly by path loss rather than multiuser interference.

\begin{figure}[!t]
\centering
\includegraphics[width=0.9\columnwidth]{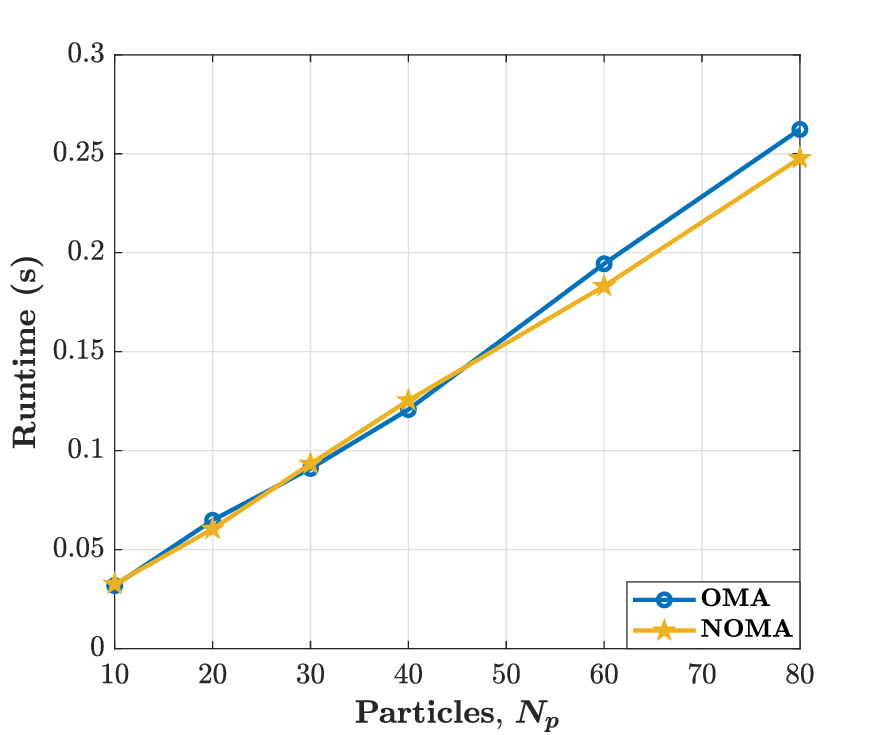}
\caption{Average runtime versus number of particles $N_p$ for OMA and NOMA.}\label{fig:runt}
\end{figure}

Fig.~\ref{fig:runt} reports wall-clock runtime versus swarm size $N_p$. The curves grow almost linearly, reflecting that each added particle incurs essentially the same amount of per-iteration complexity for channel/geometry evaluation and constraint checks. OMA and NOMA exhibit very similar slopes because their outer-loop computations are identical; the small gap arises from the inner update, where OMA uses a simple 1D search, whereas NOMA performs a slightly heavier but still low-dimensional numerical search. Absolute times shift with the number of PSO iterations and the preset resolution, but the near-linear dependence on $N_p$ remains.

\vspace{-2mm}
\section{Conclusions}\label{sec:conclusion}
In this work, we developed a coverage-centric framework for FIRES under both OMA and NOMA. We derived closed-form far-field bounds that expose the roles of aperture, number of active elements, ES, and phase-control errors, and we embedded these bounds in bi-level designs comprising a 1D inner search for OMA and a low-dimensional numerical search for NOMA, wrapped by a particle-swarm outer search over element positions with spacing constraints. The resulting methodology provides interpretable limits and practical algorithms whose cost scales nearly linearly with the swarm size. Numerical results consistently showed that FIRES enlarges the total coverage region compared with a same-size STAR-RIS that lacks position agility, and that NOMA offers additional gains when SIC is feasible. The trends versus average SNR, element count, phase-quantization resolution, phase jitter, and Rician factor closely follow the proposed bounds, confirming both their accuracy and their utility for design. The optimizer converges quickly in tens of iterations, and the observed robustness to phase errors indicates that a few bits of phase control already capture most of the gain.  While the present study focuses on analytical characterization and system-level optimization under realistic implementation constraints, including discrete positioning and phase control impairments, experimental validation of a fully integrated FIRES prototype remains an important direction for future work.

\appendices

\end{document}